\documentclass[final,10pt,twoside]{IEEEtranTCOM}
\normalsize
\ifCLASSINFOpdf
\else
\fi

\usepackage{amsthm}
\usepackage{amsmath}
\usepackage{amssymb}
\usepackage{graphicx}
\usepackage{esint}
\usepackage{amsfonts}
\usepackage{cite}
\usepackage{balance}
\usepackage{caption}
\usepackage{subcaption}
\usepackage{epstopdf}
\usepackage{color}
\usepackage{tikz}

\makeatletter

\theoremstyle{plain}
\newtheorem{thm}{\protect\theoremname}

\makeatother

\providecommand{\theoremname}{Theorem}

\theoremstyle{plain}
\newtheorem{lem}{\protect\lemmaname}

\makeatother

\providecommand{\lemmaname}{Lemma}

\theoremstyle{plain}
\newtheorem{cor}{\protect\corname}

\makeatother

\providecommand{\corname}{Corollary}

\theoremstyle{plain}
\newtheorem{prop}{\protect\propname}

\makeatother

\providecommand{\propname}{Proposition}
\DeclareMathOperator{\erf}{erf}

\usepackage{pifont}% http://ctan.org/pkg/pifont
\newcommand{\cmark}{\ding{51}}%
\newcommand{\xmark}{\ding{55}}%

\begin{document}
%\markboth{Submitted to \textit{IEEE Communications Letters}}{Boulogeorgos, Chatzidiamantis, Karagiannidis, Spectrum Sensing in Multiple Primary Users Environment under Nakagami-m Fading}
\title{ Outage performance analysis of RIS-assisted UAV wireless systems under disorientation and misalignment}

\author{\vspace{-0.2cm} 
Alexandros-Apostolos A. Boulogeorgos, Senior Member, IEEE,   Angeliki Alexiou, Member, IEEE, and Marco Di Renzo, Fellow, IEEE  \vspace{-0.8cm} 

\thanks{A.-A. A. Boulogeorgos and A. Alexiou are with the of Digital Systems, University of Piraeus Piraeus 18534 Greece (e-mails: al.boulogeorgos@ieee.org, alexiou@unipi.gr).
}

\thanks{M. Di Renzo is with Universit\'e Paris-Saclay, CNRS, CentraleSup\'elec, Laboratoire des Signaux et Syst\`emes, 3 Rue Joliot-Curie, 91192 Gif-sur-Yvette, France. (marco.di-renzo@universite-paris-saclay.fr)} 

%\thanks{M.-S. Alouini is with the Department of Computer,
%	Electrical and Mathematical Sciences and Engineering (CEMSE),
%	King Abdullah University of Science and Technology (KAUST),
%	Thuwal, Makkah Province, Kingdom of Saudi Arabia, 23955-6900 (e-mail: slim.alouini@kaust.edu.sa).}
%}

%\thanks{...}

\thanks{This work has received funding from the European Unions Horizon 2020
	research and innovation programme under grant agreement No. 871464
	(ARIADNE).}

}
\maketitle	
%\vspace{-0.4cm}
\begin{abstract}
  In this paper, we analyze the performance of a reconfigurable intelligent surface (RIS)-assisted unmanned aerial vehicle (UAV) wireless system that is affected by mixture-gamma small-scale fading, stochastic disorientation and misalignment, as well as transceivers hardware imperfections. First, we statistically characterize the end-to-end channel for both cases, i.e., in the absence as well as in the presence of disorientation and misalignment, by extracting closed-form formulas for the probability density function (PDF) and the cumulative distribution function (CDF). Building on the aforementioned expressions, we extract novel closed-form expressions for the outage probability (OP) in the absence and the presence of disorientation and misalignment as well as hardware imperfections. In addition, high signal-to-noise ratio OP approximations are derived, leading to the extraction of the diversity order. Finally, an OP floor due to disorientation and misalignment is presented.           
\end{abstract}
%\vspace{-0.2cm}

\vspace{-0.5cm}
\section{Introduction}\label{S:Intro}
%\vspace{-0.2cm}

By setting the stage for a range of novel use cases, such as remote sensing, monitoring and surveillance, ad-hoc and rapid connectivity as well as network augmentation, unmanned aerial vehicles (UAVs) have been widely recognized as a key technology for the beyond the fifth generation (B5G) era~\cite{Cho2011,Gupta2016,Trotta2018,Khan2018,Sawadsitang2019,Xu2020,Pliatsios2021,Ranjha2021,Azari2021,Geraci2021}. From a communications perspective, all UAV-based applications have two common requirements: i) ultra-reliable connectivity to ensure a high quality of experience~\cite{Xiao2021,Araniti2021,Boulogeorgos2021a}, and ii) high-energy efficiency to guarantee acceptable hover time~\cite{Zeng2017,Muntaha2021,Babu2021}. These requirements can be technically translated into the need to create a favorable and reconfigurable electromagnetic environment without increasing the UAV's energy consumption. 

In response to the above need, several researchers have recently focused their attention on combining UAVs and reconfigurable intelligent surfaces (RIS)~\cite{Renzo2019,Renzo2020}. As a result of these efforts two system models have been developed. In the first, the RIS is attached to the UAV~\cite{Li2021,Mursia2021, Shafique2021, AlJarrah2021, Yao2021, Shang2021}, while in the second, the RIS is attached to a fixed position, such as a wall, and has a clear line of sight (LOS) to the RIS~\cite{Li2020a,Nguyen2021,Li2021a,Hua2021,Mu2021,Agrawal2021,Guo2021,Pan2021,Yang2020}. For both  system models, great effort have been made to optimize and analyze their performance. In~\cite{Li2021},  the decoding error probability was minimized by optimizing the UAV placement, transmission block length, RIS meta-atoms (MAs) phase shifts (PSs), and power allocation in a UAV wireless system where the RIS was attached to the UAV. In~\cite{Mursia2021}, a problem of maximization the minimum signal-to-noise ratio (SNR) in a target area was formulated and solved for the same system model as in~\cite{Li2021}. In~\cite{Shafique2021}, the performance of UAV with attached RIS wireless systems was quantified in terms of outage probability (OP), energy efficiency, and ergodic capacity, assuming that both the source (S)-UAV and UAV-destination channels experience Rician small-scale fading. Similarly, in~\cite{AlJarrah2021}, the authors analyzed the capacity of UAV with attached RIS wireless systems in Rician fading environments under imperfect phase compensation. 
In~\cite{Yao2021}, the energy efficiency of the UAV with attached RIS wireless systems  with Fisher-Snedecor $\mathcal{F}$-composite fading was evaluated.    
Finally, in~\cite{Shang2021},  the authors presented a UAV swarm-enabled airborne RIS-empowered wireless system and quantified its performance in terms of achievable data rate using Monte Carlo simulations.           
   
   For the second system model, the authors in~\cite{Li2020a} studied the problem of optimizing the joint UAV trajectory and the passive beamforming design of the RIS in order to maximize the average achievable data rate. In addition, in~\cite{Nguyen2021}, a policy to maximize the energy efficiency  was derived for RIS-assisted UAV wireless systems operating in propagation environments with Rician fading. In~\cite{Li2021a}, the authors proposed a robust joint design of the UAV trajectory, RIS passive beamforming, and legitimate transmitter power that aims to maximize the minimum achievable secrecy rate in a RIS-assisted UAV system used to provide enhanced physical layer security in a Rician fading environment. 
   In~\cite{Hua2021}, a weighted sum bit error rate minimization problem among all RIS was studied by jointly optimizing the UAV trajectory, RIS MAs PS, and RIS scheduling, assuming independent and identical Rician distributed S-RIS and RIS-UAV channels.  
   In~\cite{Mu2021}, the three-dimensional placement and transmit power of UAVs, the reflection matrix of the RIS, and the decoding order between users were jointly optimized to maximize of the sum rate of a non-orthogonal multiple access RIS-assisted UAV wireless network. Again, both the S-RIS and RIS-UAV channels were modeled as Rician-distributed random variables~(RVs). 
   
   Furthermore, in~\cite{Agrawal2021}, the performance of an interference-limited RIS-assisted UAV wireless system suffering from Nakagami-$m$ fading was evaluated in terms of coverage probability, bit and block error rates, as well as goodput. 
   In~\cite{Guo2021}, the active beamforming of the UAV, the coefficients of the RIS MAs, and the trajectory of the UAV were jointly optimized to maximize the overall secrecy rate of all legitimate users in the presence of multiple eavesdroppers in RIS-empowassistedered UAV wireless systems operating in the millimeter-wave band and exhibiting Rayleigh fading. In~\cite{Pan2021}, the authors investigated the joint optimization of UAV trajectory, RIS PS, sub-band allocation, and  power management  to maximize the minimum average achievable rate of all users in a RIS-assisted UAV wireless system operating in the terahertz (THz) band. Finally, in~\cite{Yang2020}, the OP, average bit error rate, and ergodic capacity of RIS-assisted dual-hop UAV communication systems were quantified, where the S-RIS and RIS-UAV links were  modeled as Rayleigh and mixture-Gamma (MG) RVs, respectively.        
   
   \begin{table}
   	\centering%
   	\renewcommand{\arraystretch}{1.3}
   	\caption{Comparison with the most related contributions.}
   	\label{T:Novelty}
   	\begin{tabular}{|p{0.097\linewidth}||
   			p{0.155\linewidth}|
   			p{0.155\linewidth}|
   			p{0.2\linewidth}|
   			p{0.15\linewidth}|}
   		\hline 
   		\begin{center}\textbf{Paper} \end{center}& \begin{center} \textbf{S-RIS channel} \end{center} & \begin{center} \textbf{RIS-UAV channel} \end{center} & \begin{center} \textbf{Disorientation \& misalignment} \end{center} &\begin{center}\textbf{Hardware imperfections} \end{center}\\
   		\hline\hline
   		\cite{Li2020a} & Deterministic & Deterministic & \hspace{+0.8cm}\xmark  &  \hspace{+0.6cm}\xmark \\
   		\hline
   		\cite{Nguyen2021} & Rician & Rician &  \hspace{+0.8cm}\xmark & \hspace{+0.6cm}\xmark \\
   		\hline 
   		\cite{Li2021a} & Rician & Rician &  \hspace{+0.8cm}\xmark & \hspace{+0.6cm}\xmark \\
  		\hline
  		\cite{Hua2021} & Rician & Rician &  \hspace{+0.8cm}\xmark & \hspace{+0.6cm}\xmark \\
  		\hline
  		\cite{Mu2021} & Rician & Rician &  \hspace{+0.8cm}\xmark & \hspace{+0.6cm}\xmark \\
  		\hline
  		\cite{Agrawal2021} & Nakagami-$m$ & Nakagami-$m$ & \hspace{+0.8cm}\xmark & \hspace{+0.6cm}\xmark \\
  		\hline
  		{\cite{Guo2021}} & Rayleigh & Rayleigh & \hspace{+0.8cm}\xmark & \hspace{+0.6cm}\xmark \\
  		\hline
  		\cite{Pan2021} & Deterministic & Deterministic & \hspace{+0.8cm}\xmark  &  \hspace{+0.6cm}\xmark \\
  		\hline
  		\cite{Yang2020} & Rayleigh & MG & \hspace{+0.8cm}\xmark  &  \hspace{+0.6cm}\xmark \\
  		\hline\hline
  		{Proposed} & MG & MG &  \hspace{+0.8cm}\cmark & \hspace{+0.6cm}\cmark\\
  		\hline
   	\end{tabular}
   \end{table}

\vspace{-0.4cm}
\subsection{Novelty and contribution}\vspace{-0.2cm}

As summarized in Table~\ref{T:Novelty}, no generalized channel model has yet been presented that captures the specifics of RIS-assisted UAV wireless systems and enables performance evaluation in different propagation environments.    
Likewise, most of the aforementioned contributions assume that the RIS-UAV link is not directional. As a result, the adverse effects of UAV disorientation and/or  misalignment of the RIS-UAV beam have been neglected. However, as we move toward next-generation wireless systems, both the operating frequency and thus the directionality of links are expected to increase~\cite{Farrag2021,Boulogeorgos2018,WP:Wireless_Thz_system_architecture_for_networks_beyond_5G,Boulogeorgos2021a}. Therefore, even small disorientations and/or misalignments may adversely affect the performance of the RIS-assisted UAV wireless system~\cite{Boulogeorgos2019,Boulogeorgos2020}. Aside from the joint effects of disorientation and misalignment, another important performance-limiting factor in high-frequency communications, that, to the best of the authors' knowledge, has not yet been investigated, is the effect of transceiver hardware imperfections~\cite{PhD:Boulogeorgos,A:IQSC,C:Energy_Detection_under_RF_impairments_for_CR,Boulogeorgos2020b,Boulogeorgos2018a,B:Schenk-book}. 

Motivated by the above observations, this work focuses on deriving a generalized  framework for evaluating the outage performance of RIS-assisted UAV wireless systems that takes into account the effects of various fading conditions, disorientation, misalignment, and/or hardware imperfections. Specifically, the technical contribution of the paper is summarized as follows: 
\begin{itemize}
	\item We provide a comprehensive system model that accounts for the effects of various small-scale fading conditions, the joint effect of stochastic disorientation and misalignment, as well as transceiver hardware imperfections. In contrast to previous publications, we assume independent MG fading in both S-RIS and RIS-UAV channels. 
	\item We statistically characterize the end-to-end (e2e) channel, for both cases, i.e., absence and presence of disorientation and misalignment, by extracting closed-form expressions for the probability density function (PDF) and the cumulative distribution function (CDF).
	\item  Building on the aforementioned contributions, we quantify the performance of RIS-assisted UAV wireless systems terms of OP and diversity order. In particular, we extract closed-form formulas for the OP and diversity order for the following cases: (i) both the S-transmitter and UAV-receiver are equipped with ideal RF front-ends and the RIS-UAV link experience neither disorientation nor misalignment, (ii) both the S-transmitter and UAV-receiver are equipped with ideal RF front-ends and the RIS-UAV link experience disorientation and misalignment, (iii) both the S-transmitter and UAV-receiver are equipped with non-ideal RF front-ends and the RIS-UAV link experience neither disorientation nor misalignment, and (iv) both the S-transmitter and UAV-receiver are equipped with non-ideal RF front-end and the RIS-UAV link experience disorientation and misalignment. 
	\item In addition, simplified and insightful approximations for the OP in the high-SNR regime are provided for all cases~studied. 
	\item Finally, an OP floor is extracted due to the effects of disorientation and misalignment.   
\end{itemize}

\vspace{-0.5cm}
\subsection{Organization and notations}\vspace{-0.2cm}

%\begin{figure}
%	\centering
%	\scalebox{0.65}{\input{images/organization.tex}}
%	\caption{The organization of the paper at a glance.}
%	\label{Fig:organization}
%\end{figure}
The remainder of the paper is organized as follows: Section~\ref{S:SM} describes the system and channel models of the investigated RIS-assisted UAV wireless system. The statistical characterization of the e2e channel is presented in Section~\ref{S:E2eChannel}. Closed-form expressions and approximations for the OP and diversity order are documented in Section~\ref{S:PA}. Numerical results that validate the theoretical framework and provide insight into the effects of small-scale fading, disorientation and misalignment, as well as transceivers hardware imperfections are presented in Section~\ref{S:Results}. Finally, concluding remarks and key observations are made in Section~\ref{S:Conclusions}. 
%Figure~\ref{Fig:organization} illustrates  the structure of this paper in detail.                           

\subsubsection*{Notations} 
Unless otherwise stated, lower bold letter stands for vectors. 
The absolute value and the exponential function are respectively denoted by $|\cdot|$,  and $\exp\left(\cdot\right)$.
  The $n-$th power and the square root of $x$ are respectively represented by $x^n$ and $\sqrt{x}$.  
 $\Pr\left(\mathcal{A}\right)$ denotes the probability for the event $\mathcal{A}$ to be valid. Likewise, $\cos\left(x\right)$ gives the cosine of $x$, while $\sin\left(x\right)$ returns the sine of $x$. Moreover, $\csc\left(x\right)$ stands for the cosecant of $x$. The error-function is represented by $\erf\left(\cdot\right)$~\cite[eq. (8.250/1)]{B:Gra_Ryz_Book}.
The modified Bessel function of the second kind of order $n$ is denoted as~$\mathrm{K}_n(\cdot)$~\cite[Eq. (8.407/1)]{B:Gra_Ryz_Book}. 
The  Gamma~\cite[Eq. (8.310)]{B:Gra_Ryz_Book} function is  denoted by  $\Gamma\left(\cdot\right)$. Finally, $\,_pF_q\left(a_1, a_2, \cdots, a_p; b_1, b_2, \cdots, b_q; x\right)$ and $G_{p, q}^{m, n}\left(x\left| \begin{array}{c} a_1, a_2, \cdots, a_{p} \\ b_{1}, b_2, \cdots, b_q\end{array}\right.\right)$  respectively stand for the generalized hypergeometric function~\cite[Eq. (9.111)]{B:Gra_Ryz_Book} and the Meijer G-function~\cite[Eq. (9.301)]{B:Gra_Ryz_Book}.

\vspace{-0.5cm}
\section{System model}\label{S:SM}\vspace{-0.2cm}

\begin{figure}
	\centering
	\scalebox{0.38}{\input{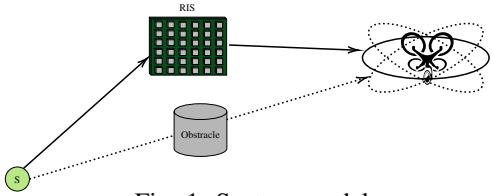}}
	\vspace{-0.25cm}
	\caption{System model}
	\label{Fig:SM}
	\vspace{-0.2cm}
\end{figure}

\begin{figure}
	\vspace{-0.2cm}
	\centering
	\scalebox{0.38}{\input{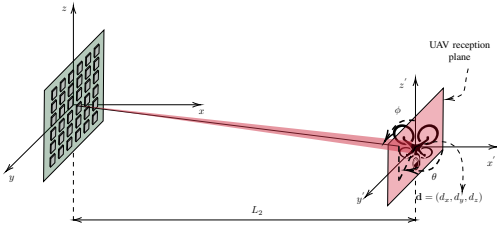}}
	\vspace{-0.2cm}
	\caption{The relative position and orientation of the RIS and UAV in the considered coordinate systems.}
	\label{Fig:Coordinates}
	\vspace{-0.4cm}
\end{figure}

As illustrated in Fig.~\ref{Fig:SM}, we consider an RIS-assisted UAV wireless system in which the S communicates with a UAV via a RIS. We assume that no direct link can be established between S and the UAV, due to the presence of blockage. It is assumed that the UAV is in a hover state, where, in practice, both the position and orientation of the UAV are not completely fixed.  It is also assumed that both the S and the UAV are equipped with single antennas, while the RIS consists of $N$ MAs. 
To determine the relative position and orientation of the RIS and the UAV, we define two three-dimensional Cartesian coordinate systems, as demonstrated in Fig.~\ref{Fig:Coordinates}. The RIS is assumed to be at the origin of the Cartesian coordinate system $(x, y, z)$, i.e., at position $(0,0,0)$. At a specific timeslot, it is assumed that the UAV is located at position
%\begin{align} 
	$\mathbf{d}_{\epsilon}=\left(d_x+\epsilon_x, d_y+\epsilon_y, d_z+\epsilon_z\right)$,
%\end{align} 
	with respect to the $(x, y, z)$ coordinate system, where 
	%\begin{align}
		$\mathbf{d}=\left(d_x, d_y, d_z\right)$
	%\end{align} 
	denotes the mean of the random vector $\mathbf{d}_{\epsilon}$
	and 
	%\begin{align}
	$\boldsymbol{\epsilon}=\left(\epsilon_x,\epsilon_y,\epsilon_z\right)$ 
	%\end{align}
	are independent and identical zero-mean Gaussian distributed RVs with variance $\sigma_p^2$. For a given $\mathbf{d}_{\epsilon}$, a second Cartesian coordinate system $(x^{'}, y^{'}, z^{'})$, for which $\mathbf{d}_{\epsilon}$ is at the origin and the axes $x^{'}$, $y^{'}$, and $z^{'}$ are parallel with the axes $x$, $y$, and $z$, respectively, is considered. Let $\theta_\epsilon\in[0, 2\pi]$ be the angle between the axis $x^{'}$ and the projection of the beam vector onto the $x^{'}-y^{'}$ plane. Similarly,  $\phi_{\epsilon}\in[0,\pi]$ represents the angle between the $z^{'}$ axis and the beam vector. Notice that both $\theta_\epsilon$ and $\phi_{\epsilon}$ are randomly distributed variables that can be respectively expressed~as
%\begin{align}
	$\theta_\epsilon = \theta + \epsilon_{\theta}$
%\end{align}       
and 
%\begin{align}
	$\phi_\epsilon = \phi + \epsilon_{\phi}$,
%\end{align}
where $\theta$ and $\phi$  denote the mean of $\theta_\epsilon$ and $\phi_\epsilon$, respectively, while $\epsilon_{\theta}$ and $\epsilon_{\phi}$ are zero-mean RVs with variance $\sigma_o^2$.

By $h_i$ and $g_i$ we denote the complex coefficients of the S to $i-$th MA and $i-$th MA to UAV baseband equivalent fading channels, respectively. It is assumed that all the fading channels considered are independent, identical, slowly varying, and flat. Finally, we use $\theta_i$ to denote the response of the $i-$th MA. As a result, the baseband equivalent received signal at D can be obtained~as
\begin{align}\vspace{-0.5cm}
	r \hspace{-0.15cm} = \hspace{-0.15cm} \left\{\hspace{-0.2cm}\begin{array}{l l} 
		h_l\, A\,  \left( s+ \eta_s\right) + \eta_d + w, & \hspace{-0.25cm} \text{without disorientation} \\ & \hspace{-0.25cm} \text{or misalignment}\\
		& \\
		h_l\, h_g\, A\,  \left( s+ \eta_s\right) + \eta_d + w, & \hspace{-0.25cm} \text{with disorientation} \\ & \hspace{-0.25cm} \text{and misalignment}\end{array}\right.
	\label{Eq:r}
\end{align}
where  $s$ represents the S transmission symbol, while $w$ stands for the additive white Gaussian noise (AWGN), which is modeled as a zero-mean complex Gaussian (ZMCG) RV with variance equal to $\sigma_w^2$. Moreover, $h_l$ denotes the e2e spreading-loss coefficient and can be written~as \vspace{-0.2cm}
\begin{align} 
	h_l = h_{l,1}\,h_{l,2},
\end{align}
\vspace{-0.3cm}where 
\begin{align}
	h_{l,i} = l_i\,L_{i}^{-n_i/2},\text{ with } i\in\{1, 2\},
	\label{Eq:h_l_i}
\end{align}
\vspace{-0.0cm}is the spreading loss coefficient of the S-RIS ($i=1$) and RIS-UAV ($i=2$) links respectively. In~\eqref{Eq:h_l_i}, $l_1$ and $l_2$ respectively denote the square root of the reference reception power, while $n_1$ and $n_2$ stand for the path-loss exponent of the S-RIS and RIS-UAV links, respectively. The transmission distances of the S-RIS and RIS-D links are denoted by $L_1$ and $L_2$, respectively.   

In addition, $h_g$ represents the geometric-loss caused by the disorientation of the UAV and the misalignment of the beam between the RIS and the UAV. According to~\cite{Najafi2018}, the PDF of $h_g$ can be expressed~as\vspace{-0.3cm}
\begin{align}
	f_{h_g}(x) = \frac{\rho}{B_o} \left(\frac{x}{B_o}\right)^{\zeta-1}, \text{ with } 0\leq x \leq B_o.
	\label{Eq:f_h_g}
\end{align}
In~\eqref{Eq:f_h_g}, \vspace{-0.3cm}
\begin{align}
	B_o = \erf\left(v_{\min}\right) \erf\left(v_{\max}\right),
\end{align}
where \vspace{-0.3cm}
\begin{align}
	v_{i} = \frac{\alpha}{w(L_2)} \sqrt{\frac{\pi}{2\rho_{i}}},\text{ with } i\in\{\min, \max\}.
	\label{Eq:v_i}
\end{align}
In~\eqref{Eq:v_i}, 
%\begin{align}
	$\rho_{\min} = \frac{2}{\rho_y+\rho_z + \sqrt{\left(\rho_y-\rho_z\right)^2 + 4 \rho_{zy}^2}}$
%\end{align}
and
%\begin{align}
	$\rho_{\max} = \frac{2}{\rho_y+\rho_z - \sqrt{\left(\rho_y-\rho_z\right)^2 + 4 \rho_{zy}^2}}$
%\end{align}
where 
%\begin{align}
	$\rho_y = \cos^2\left(\phi\right) + \sin^2\left(\phi\right) \cos^2\left(\theta\right)$, 
	%\\
	$\rho_z = \sin^2\left(\phi\right)$
%\end{align}
and
%\begin{align}
$	\rho_{yz} = - \cos\left(\phi\right) \sin\left(\phi\right) \sin\left(\theta\right).$	
%\end{align}
Moreover, $\alpha$ is the radius of the UAV antenna effective area, while $w(L_2)$ is  the RIS beamwidth at distance $L_2$ and can be expressed~as
%\begin{align}
	$w\left(L_2\right) = w_o \sqrt{1+\left(1+\frac{2 w_o^2}{\rho^2\left(L_2\right)} \right)\left( \frac{c L_2}{\pi f w_o^2} \right)^2},$ 
%\end{align}    
with $w_o$ being the beam-waist radius, $c$ denoting the speed of light, $f$ standing for the transmission frequency, and $\rho\left(L_2\right)$ representing the coherence length. Of note, the coherence length can be evaluated~as
%\begin{align}
	$\rho\left(L_2\right) = \left(0.55\, C_n^2\, k^2 \, L_2\right)^{-3/5},$
%\end{align}
where $C_n^2$ stands for the index of refraction structure parameter and 
%\begin{align}
	$k=\frac{2\pi f}{c},$
%\end{align} 
	is the wave-number.  
In~\eqref{Eq:f_h_g},
%\begin{align}
	$\zeta = \frac{k_m \left(w(L_2)\right)^2}{4\sigma_p^2+4 d_x^2 \sigma_0^2}.$
%\end{align}
with
%\begin{align}
	$k_m = \frac{k_{\min}+k_{\max}}{2},$
%\end{align}
where 
%\begin{align}
	$k_{i}=\frac{\sqrt{\pi}\rho_{i} \erf\left(v_{i}\right)}{2v_{i}\exp\left(-v_{i}^2\right)}, \text{ with } i\in\left\{\min, \max\right\}.$
%\end{align}

Finally, \vspace{-0.3cm}
\begin{align}
	A= \sum_{i=1}^{N} h_i \theta_i g_i, 
	\label{Eq:A}
\end{align}     
where $h_i$ and $g_i$ are the S-$i-$th MA and $i-$th MA-D channel coefficients, which can be expressed as %\vspace{-0.0cm}
\begin{align}
	h_i = \left|h_i\right| \exp(j\phi_{h_i})
%\end{align}  
\text{ and }
%\begin{align}
	g_i = \left|g_i\right| \exp(j\phi_{g_i}),
\end{align}
with $\phi_{h_i}$ and $\phi_{g_i}$ respectively being the phases of $h_i$ and $g_i$. The envelops of $h_i$ and $g_i$ are assumed to be independent MG random variables\footnote{Note that the MG distribution has been proven to be able to very accurately approximate a number of widely used distributions, such as Rayleigh, Rice, Nakagami-$m$, Gamma, generalized Gamma, etc.} with PDFs that can be respectively expressed~as \vspace{-0.3cm}
\begin{align}
	f_{h_i}(x) = \sum_{m=1}^{M} 2 a^{(1)}_{m} x^{2 b^{(1)}_m-1}\exp\left(-c_1 x^2\right)
	\label{Eq:f_h_i}
\end{align}
and \vspace{-0.4cm}
\begin{align}
	f_{g_i}(x) = \sum_{k=1}^{K} 2 a^{(2)}_{k} x^{2 b^{(2)}_k-1}\exp\left(-c_2 x^2\right),
	\label{Eq:f_g_i}
\end{align} 
 where $M$ and $K$ are the numbers of terms for the approximation of the PDF of $\left|h_i\right|$ and $\left|g_i\right|$, respectively, while $a^{(1)}_{m}$, $b^{(1)}_m$ and $c_1$ with $m\in[1, M]$ are parameters of the $m-$th term of~\eqref{Eq:f_h_i}. Similarly, $a^{(2)}_{k}$, $b^{(2)}_k$ and $c_2$ with $k\in[1, K]$ are parameters of the $k-$th term of~\eqref{Eq:f_g_i}.

Moreover, in~\eqref{Eq:A}, $\theta_i$ stands for the $i-$th MA response and can be further written~as \vspace{-0.3cm}
\begin{align}
	\theta_i = \left|\theta_i\right| \exp\left(j\phi_i\right), 
	\label{Eq:theta_i} \vspace{-0.3cm}
\end{align} 
with $\left|\theta_i\right|$ and $\phi_i$ respectively representing the $i-$th MA response gain and the PS applied by the $i-$th MA of the RIS. Without loss of generality, we assume that $\left|\theta_i\right|=1$. Based on~\cite{Asadchy2016}, this is considered a realistic assumption. As reported in several works including~\cite{Boulogeorgos2020a,Basar2019}, the optimal PS for the $i-$th MA~is  \vspace{-0.3cm}
\begin{align}
	\phi_i = -\phi_{h_i} - \phi_{g_i}. 
	\label{Eq:phi_i_optimal}
\end{align} 
By applying~\eqref{Eq:phi_i_optimal} to~\eqref{Eq:A}, we get  \vspace{-0.2cm}
\begin{align}
	A=\sum_{i=1}^{N} \left|h_i\right| \left|g_i\right|. 
	\label{Eq:A_final}
\end{align}

In~\eqref{Eq:r}, $\eta_s$ and $\eta_d$ models respectively for the impact of S and UAV RF chains hardware imperfections. According to~\cite{Boulogeorgos2016,C:Energy_Detection_under_RF_impairments_for_CR,A:LC_CR_vs_SS}, for a given e2e channel realization,  $\eta_s$ and $\eta_d$ can be modeled as two independent ZMCG RVs with variances that can be respectively expressed~as  \vspace{-0.3cm}
\begin{align}
	\sigma_s^2=\kappa_s^2\,P_s
\end{align}
and  \vspace{-0.3cm}
\begin{align}
	\sigma_d^2 = \kappa_d^2\, h_l^2\, h_g^2\, A^2\, P_s,
\end{align}
where $\kappa_s$ and $\kappa_d$ are respectively the S transmitter and UAV receiver error vector magnitudes, while $P_s$ stands for the average transmitted power. According to~\cite{Boulogeorgos2019,Boulogeorgos2020,Boulogeorgos2020b}, $\kappa_s$ and $\kappa_d$ in high-frequency systems, such as millimeter wave and terahertz (THz), are in the range of $[0.07, 0.3]$. Finally, for the special case in which both the S and UAV are equipped with ideal transceivers, $\kappa_s=\kappa_d=0$~\cite{Bjoernson2013}.  

 \vspace{-0.3cm}
\section{E2e channel statistical characterization}\label{S:E2eChannel}  \vspace{-0.2cm}

This section focuses on statistically characterizing the distribution of the e2e equivalent channel, i.e.,  %\vspace{-0.3cm}
\begin{align}
	A_{\mathrm{e2e}} = h_g\, A.
\end{align} 
In this direction, the following theorem returns novel closed-form expressions for the PDF and CDF of $A$. 

\begin{thm}\label{L:PDF_CDF_A}
	The PDF and CDF of $A$ can be respectively evaluated~as  \vspace{-0.3cm}
	\begin{align}
		f_{A}(x) = \frac{4 \Xi^{k_A + m_A}}{\Gamma\left(k_A\right)\Gamma\left(m_A\right)} x^{k_A+m_A-1} \mathrm{K}_{k_A-m_A}\left(2\Xi x\right)
		\label{Eq:f_A}
	\end{align}
	and  \vspace{-0.3cm}
	\begin{align}
		F_{A}(x) = \frac{1}{\Gamma\left(k_A\right)\Gamma\left(m_A\right)} \mathrm{G}_{1, 3}^{2, 1}\left(\Xi^2 x^2\left|\begin{array}{c} 1\\ k_A, m_A, 0 \end{array}\right.\right),
		\label{Eq:F_A}
	\end{align}
	where  \vspace{-0.3cm}
	\begin{align}
		k_A = - \frac{b_A}{2 a_A} + \frac{\sqrt{b_A^2 - 4 a_A c_A}}{2 a_A},
		\label{Eq:k_A}
	\end{align}  \vspace{-0.5cm}
	\begin{align}
		m_A = - \frac{b_A}{2 a_A} - \frac{\sqrt{b_A^2 - 4 a_A c_A}}{2 a_A}
	\end{align}
	and  \vspace{-0.3cm}
	\begin{align}
		\Xi = \sqrt{\frac{k_A m_A}{\Omega_A}}.
		\label{Eq:Xi_A}
	\end{align}
In~\eqref{Eq:k_A}--\eqref{Eq:Xi_A},
\begin{align}
	a_A \hspace{-0.15cm}&=\hspace{-0.15cm} \mu_{A}\left(6\right) \mu_{A}\left(2\right) + \left(\mu_{A}\left(2\right)\right)^2 \mu_{A}\left(4\right) - 2\left(\mu_{A}\left(4\right)\right)^2,
	\\
	b_A\hspace{-0.15cm}&=\hspace{-0.15cm} \mu_{A}\left(6\right) \mu_{A}\left(2\right)- 4 \left(\mu_{A}\left(4\right)\right)^2 + 3 \left(\mu_{A}\left(2\right)\right)^2 \mu_{A}\left(4\right),
\\
	c_A &= 2 \left(\mu_{A}\left(2\right)\right)^2 \mu_{A}\left(4\right)
\end{align}
and \vspace{-0.3cm}
\begin{align}
	\Omega_A = \mu_A\left(2\right),
\end{align}
where \vspace{-0.3cm}
\begin{align}
	\mu_{A}\left(l\right) = \sum_{l_1=0}^{l}&\sum_{l_2=0}^{l_1}\cdots\sum_{l_{N-1}=0}^{l_{N-2}}
	\left(\begin{array}{c}l\\l_1\end{array}\right) \left(\begin{array}{c}l_1\\l_2\end{array}\right)
	\cdots 
	\left(\begin{array}{c}l_{N-2}\\l_{N-1}\end{array}\right)
	\nonumber \\ 
	& \hspace{-0.7cm} \times \mu_{\chi_1}\left(l-l_1\right) \mu_{\chi_2}\left(l_1-l_2\right) 
	\cdots \mu_{\chi_{N-1}}\left(l_{N-1}\right)
	\label{Eq:mu_A}
\end{align}
and \vspace{-0.3cm}
\begin{align}
	\mu_{\chi_i}(l) = \sum_{m=1}^{M}&\sum_{k=1}^{K}
	a_m^{(1)} a_k^{(2)} \left(\frac{c_1}{c_2}\right)^{-\frac{b_m^{(1)}-b_k^{(2)}}{2}} \left(c_1 c_2\right)^{-\frac{b_m^{(1)}+b_k^{(2)}+n}{2}}
	\nonumber \\ & \times
	\Gamma\left(b_{m}^{(1)}+\frac{n}{2}\right) \Gamma\left(b_{k}^{(2)}+\frac{n}{2}\right)
	\label{Eq:mu_x}
\end{align}
\end{thm}
\begin{IEEEproof}
	For brevity, the proof of Lemma~\ref{L:PDF_CDF_A} is given in Appendix A. 
\end{IEEEproof}

The following lemma returns an alternative formula for the CDF of $A$ for the special case in which $k_A\neq j m_A$ and $m_A \neq j k_A$ with $j=1, 2, 3, \cdots$. 
\begin{lem}
	For the special case in which $k_A\neq j\,\, m_A$  and $m_A \neq j k_A$ with $j=1, 2, 3, \cdots$, the CDF of $A$ can be rewritten as in~\eqref{Eq:F_A_s}, given at the top of the next page.
	\begin{figure*}  
	\begin{align}
		F_{A}^{(s)}(x) &=  \frac{\pi \csc\left(\left(k_A-m_A\right)\pi\right) \Xi^{2m_A} }{\Gamma\left(k_A\right)  \Gamma\left(m_A+1\right)\Gamma\left(1-k_A+m_A\right)} x^{2m_A}\,_1F_2\left(m_A;1+m_A, 1-k_A+m_A; \Xi^2 x^2\right)
		\nonumber \\ &  
		- \frac{\pi \csc\left(\left(k_A-m_A\right)\pi\right) \Xi^{2k_A} }{\Gamma\left(m_A\right)  \Gamma\left(k_A+1\right)\Gamma\left(1+k_A-m_A\right)} x^{2k_A}
		\,_1F_2\left(k_A;1+k_A, 1+k_A-m_A; \Xi^2 x^2\right)
		\label{Eq:F_A_s}
	\end{align}
	\vspace{-0.3cm}
	\hrulefill
	\end{figure*}
\end{lem}
\begin{IEEEproof}
	By applying~\cite[Eq. (9.303)]{B:Gra_Ryz_Book} in~\eqref{Eq:F_A}, we obtain~\eqref{Eq:F_A_s}. This concludes the proof.
\end{IEEEproof}

The following theorem returns the PDF and CDF of $A_{\mathrm{e2e}}$. 
\begin{thm}\label{Th:PDF_CDF_Ae2e}
	The PDF and CDF of $A_{\mathrm{e2e}}$ can be respectively obtained~as 	\vspace{-0.3cm}
	\begin{align}
		f_{A_{e2e}}&\left(x\right) = -\frac{\zeta}{\Gamma\left(k_A\right)\Gamma\left(m_A\right)} x^{-1}
		\nonumber \\ & \hspace{-0.6cm} \times
		\left( 
		\mathrm{G}_{4,2}^{1,3}\left(\left.\begin{array}{c} 1-k_A, 1-m_A, \frac{1-\zeta}{2}, 1 \\ 0, \frac{1-\zeta}{2}\end{array}\right|  \frac{B_o^2}{\Xi^2 x^2} \right)
		\right.
		\nonumber \\ & \hspace{-0.6cm}
		\left.
		-\frac{\zeta}{2}
		\mathrm{G}_{5,3}^{1,4}\left(\left.\begin{array}{c} 1-k_A, 1-m_A, \frac{1-\zeta}{2}, \frac{2-\zeta}{2},1 \\ 0, \frac{1-\zeta}{2}, -\frac{\zeta}{2} \end{array}\right| \frac{B_o^2}{\Xi^2 x^2} \right)
		\right)
		\label{Eq:f_A_e2e_final}
	\end{align}
and 	\vspace{-0.3cm}
	\begin{align}
		F_{A_{e2e}}&(x)= \frac{\zeta}{2\Gamma\left(k_A\right)\Gamma\left(m_A\right)}
		\nonumber \\ & \hspace{-0.6cm}\times \mathrm{G}_{5,3}^{1,4}\left(\left.\begin{array}{c}1-k_A, 1-m_A, \frac{1-\zeta}{2}, \frac{2-\zeta}{2},1 \\ 0, \frac{1-\zeta}{2}, - \frac{\zeta}{2}\end{array}\right| \frac{B_o^2}{\Xi^2 x^2}\right).
		\label{Eq:F_A_e2e_final_general}
	\end{align}
\end{thm}
\begin{IEEEproof}
	For brevity, the proof of Theorem~\ref{Th:PDF_CDF_Ae2e} is provided in Appendix~B. 
\end{IEEEproof}

The following lemma returns an alternative formula for the CDF of $A_{e2e}$ for the special case in which $k_A\neq j m_A$ with $j=1, 2, 3, \cdots$.  
\begin{lem}\label{l:special_case_CDF}
	For the special case in which $k_A\neq j\,\, m_A$ with $j=1, 2, 3, \cdots$, the CDF of $A_{e2e}$ can be rewritten as in~\eqref{Eq:F_A_2e2_final_special}, given at the top of the next page.  
	\begin{figure*}
		\begin{align}
			F_{A_{e2e}}^{(s)}(x) &=  \frac{\Xi^\zeta \Gamma\left(k_A-\frac{\zeta}{2}\right) \Gamma\left(m_A-\frac{\zeta}{2}\right)}{B_o^{\zeta} \Gamma\left(k_A\right) \Gamma\left(m_A\right)}
			+ \frac{\rho\, \Xi^{2k_A}}{k_A \zeta B_o^{2\zeta } }\frac{\Gamma\left(-k_A+m_A\right)}{\Gamma\left(k_A\right)\Gamma\left(m_A\right)}  x^{2k_A}\, \,_1F_2\left(k_A; 1+k_A, 1+k_A-m_A; \frac{\Xi^2}{B_o^2}x^2\right)
			\nonumber \\ & 
			-  \frac{2 \Xi^{2k_A}}{ \left(2k_A-\rho\right) B_o^{2\zeta}} \frac{\Gamma\left(-k_A+m_A\right)}{\Gamma\left(k_A\right)\Gamma\left(m_A\right)}  x^{2k_A}\, \,_1F_2\left(k_A-\frac{\zeta}{2}; 1+k_A-m_A, 1+k_A-\frac{\zeta}{2}; \frac{\Xi^2}{B_o^2} x^2\right)
			\nonumber \\ & 
			+\frac{2\rho  \Xi^{2m_A}}{m_A\zeta B_o^{2m_A}}\frac{\Gamma\left(k_A-m_A\right)}{\Gamma\left(k_A\right)\Gamma\left(m_A\right)}  x^{2m_A}\,\,_1F_2\left(m_A;1+m_A, 1-k_A+m_A; \frac{\Xi^2 x^2}{B_o^2}\right)
			\nonumber \\ &
			-\frac{2\rho\Xi^{2m_A}}{\left(2m_A-\zeta\right)\zeta B_o^{2m_A}}\frac{\Gamma\left(k_A-m_A\right)}{\Gamma\left(k_A\right)\Gamma\left(m_A\right)}  x^{2m_A} \, \,_1F_2\left(m_A - \frac{\zeta}{2}; 1-k_A+m_A, 1+m_A-\frac{\zeta}{2}; \frac{\Xi^2 x^2}{B_o^2}\right) 
			\label{Eq:F_A_2e2_final_special}
		\end{align}
		\vspace{-0.3cm}
		\hrulefill
	\end{figure*}
\end{lem}
\begin{IEEEproof}
	By applying~\cite[Eq. (9.303)]{B:Gra_Ryz_Book} in~\eqref{Eq:F_A_e2e}, we obtain~\eqref{Eq:F_A_2e2_final_special}. This concludes the proof.  
\end{IEEEproof}

\vspace{-0.3cm}
\section{Performance analysis}\label{S:PA} \vspace{-0.1cm}

This section is focused on presenting the theoretical framework that quantifies the outage performance and diversity order of the RIS-assisted UAV wireless system in the absence and presence of transceiver hardware imperfections. Specifically, in Section~\ref{SS:Ideal_RF_FE}, closed-form expressions for the OP and diversity order in the case that both the S and UAV are equipped with ideal RF front-ends are presented, while, in Section~\ref{SS:NIdeal_RF_FE}, the corresponding expressions for the case in which both the S and UAV suffer from hardware imperfections are extracted.  

\vspace{-0.3cm}
\subsection{Ideal RF front-end}\label{SS:Ideal_RF_FE}
The goal of this section is to present a theoretical framework that is suitable for characterizing the outage performance of RIS-assisted UAV wireless systems in the absence of transceivers hardware imperfections. In this direction, Section~\ref{SS:SNR} focuses on the extraction of the instantaneous SNR, while Section~\ref{SS:OP_woH} returns the corresponding outage performance analysis framework. Finally, Section~\ref{SS:DO} presents the diversity order.     

\subsubsection{Instantaneous SNR}\label{SS:SNR}
For the case in which both the S and UAV transceivers experience no hardware imperfections, the SNR at the UAV can be obtained~as
\begin{align}
	\gamma_u^{\text{id}}\hspace{-0.1cm} =\hspace{-0.1cm} \left\{\hspace{-0.2cm} \begin{array}{l} \gamma A^2, \text{ } \text{ without disorientation or misalignment }\\
	\gamma A_{e2e}^2, \text{ with disorientation and misalignment} \end{array}\right. 
\label{Eq:gamma_u_ideal}
\end{align}
where
%\begin{align}
	$\gamma = \frac{h_l^2 P_s}{\sigma_w^2}.$
%\end{align}

\subsubsection{OP}\label{SS:OP_woH}
In this section, closed-form expressions and approximations for the OP for the case in which neither the S transmitter nor the UAV receiver experience RF front-end imperfections are provided.   

\underline{Without disorientation and misalignment}: The following proposition returns the OP for the case in which both the S and UAV transceivers are equipped with ideal RF-front and the RIS-assisted UAV wireless system experience neither disorientation nor misalignment. 
\begin{prop}
	For the case in which both the S and UAV transceivers are equipped with ideal RF front-ends and the RIS-assisted UAV wireless system suffers from neither disorientation nor misalignment, the OP can be obtained~as 	\vspace{-0.2cm}
	\begin{align}
		P_o^{\text{id-wo}}(\gamma_{\mathrm{th}}) = \frac{\mathrm{G}_{1, 3}^{2, 1}\left(\Xi^2 \frac{\gamma_{\mathrm{th}}}{\gamma}\left|\begin{array}{c} 1\\ k_A, m_A, 0 \end{array}\right.\right)}{\Gamma\left(k_A\right)\Gamma\left(m_A\right)},
		\label{Eq:Po_id_wo}
	\end{align}
where $\gamma_{\mathrm{th}}$ stands for the SNR threshold. 
\end{prop} 
\begin{IEEEproof}
	In the case of ideal RF front-end at both the S and UAV transceivers, the OP can be defined~as 	\vspace{-0.2cm}
	\begin{align}
			P_o^{\text{id-wo}}(\gamma_{\mathrm{th}}) = \Pr\left(\gamma_u^{\text{id}} \leq \gamma_{\mathrm{th}}\right),
			\label{Eq:OP_id_wo_definition}
	\end{align}
which, by applying~\eqref{Eq:gamma_u_ideal} can be rewritten~as 	\vspace{-0.2cm}
	\begin{align}
		P_o^{\text{id-wo}}(\gamma_{\mathrm{th}}) = \Pr\left( A \leq \sqrt{\frac{\gamma_{\mathrm{th}}}{\gamma}}\right), 
	\end{align}
or equivalently 	\vspace{-0.2cm}
	\begin{align}
		P_o^{\text{id-wo}}(\gamma_{\mathrm{th}}) = F_{A}\left(\sqrt{\frac{\gamma_{\mathrm{th}}}{\gamma}}\right).
		\label{Eq:Po_id_wo_s2}
	\end{align}
Finally, by applying~\eqref{Eq:F_A} in~\eqref{Eq:Po_id_wo_s2}, we get~\eqref{Eq:Po_id_wo}. This concludes the proof. 
\end{IEEEproof}

For the special case in which no disorientation and misalignment is experienced and $k_A\neq j\,m_A$ as well as $m_A\neq j\,k_A$ with $j=1,2,\cdots$, the OP can be calculated as in~\eqref{Eq:P_o_s_id_wo}, given at the top of the next page.
\begin{figure*}
	\begin{align}
		P_{o,s}^{\text{id-wo}}(\gamma_{\mathrm{th}}) &= \frac{\pi \csc\left(\left(k_A-m_A\right)\pi\right) \Xi^{2m_A} }{\Gamma\left(k_A\right)  \Gamma\left(m_A+1\right)\Gamma\left(1-k_A+m_A\right)} \left(\frac{\gamma_{\mathrm{th}}}{\gamma}\right)^{m_A}\,_1F_2\left(m_A;1+m_A, 1-k_A+m_A; \Xi^2 \frac{\gamma_{\mathrm{th}}}{\gamma}\right)
		\nonumber \\ &  
		- \frac{\pi \csc\left(\left(k_A-m_A\right)\pi\right) \Xi^{2k_A} }{\Gamma\left(m_A\right)  \Gamma\left(k_A+1\right)\Gamma\left(1+k_A-m_A\right)} \left(\frac{\gamma_{\mathrm{th}}}{\gamma}\right)^{k_A}
		\,_1F_2\left(k_A;1+k_A, 1+k_A-m_A; \Xi^2 \frac{\gamma_{\mathrm{th}}}{\gamma}\right)
		\label{Eq:P_o_s_id_wo}
	\end{align}
	\vspace{-0.3cm}
	\hrulefill
\end{figure*} 
For this case, the following lemma returns a high-SNR OP approximation. 
\begin{lem}
	In the high-SNR regime, the OP can be approximated~as 	\vspace{-0.2cm}
	\begin{align}
		P_{o,s,\infty}^{\text{id-wo}}(\gamma_{\mathrm{th}}) &\approx \frac{\pi \csc\left(\left(k_A-m_A\right)\pi\right) \Xi^{2m_A} }{\Gamma\left(k_A\right)  \Gamma\left(m_A+1\right)\Gamma\left(1-k_A+m_A\right)} \left(\frac{\gamma_{\mathrm{th}}}{\gamma}\right)^{m_A}
		\nonumber \\ &  \hspace{-1.3cm}
		- \frac{\pi \csc\left(\left(k_A-m_A\right)\pi\right) \Xi^{2k_A} }{\Gamma\left(m_A\right)  \Gamma\left(k_A+1\right)\Gamma\left(1+k_A-m_A\right)} \left(\frac{\gamma_{\mathrm{th}}}{\gamma}\right)^{k_A}.
		\label{Eq:P_o_s_id_wo_approx}
	\end{align}
\end{lem}
\begin{IEEEproof}
	In the high-SNR regime, i.e. for $\gamma\to\infty$, $z\to 0$, with $z=\frac{\gamma_{\mathrm{th}}}{\gamma}$, it holds that  	\vspace{-0.2cm}
	\begin{align}
		\lim_{z\to 0} \,_1F_2\left(s_1;t_1, t_2; z\right) = 1. 
		\label{Eq:limit_z_0}
	\end{align}
	By applying~\eqref{Eq:limit_z_0} into~\eqref{Eq:P_o_s_id_wo}, we obtain~\eqref{Eq:P_o_s_id_wo_approx}. This concludes the~proof. 
\end{IEEEproof}

\underline{With disorientation and misalignment}: The following proposition returns the OP for the case in which both the S and UAV transceivers are equipped with ideal RF-front and the RIS-assisted UAV wireless system suffers from disorientation and misalignment. 
\begin{prop}
	For the case in which both the S and UAV transceivers are equipped with ideal RF-front and the RIS-assisted UAV wireless system suffers from disorientation and misalignment, the OP can be expressed~as \vspace{-0.2cm}
	\begin{align}
		P_{o}^{\text{id-w}}&(\gamma_{\mathrm{th}})= \frac{\zeta}{2\Gamma\left(k_A\right)\Gamma\left(m_A\right)}
		\nonumber \\ & \hspace{-0.6cm}\times \mathrm{G}_{5,3}^{1,4}\left(\left.\begin{array}{c}1-k_A, 1-m_A, \frac{1-\zeta}{2}, \frac{2-\zeta}{2},1 \\ 0, \frac{1-\zeta}{2}, - \frac{\zeta}{2}\end{array}\right| \frac{B_o^2}{\Xi^2} \frac{\gamma}{\gamma_{\mathrm{th}}}\right).
		\label{Eq:P_o_id_w}
	\end{align}
\end{prop}
\begin{IEEEproof}
	From \eqref{Eq:OP_id_wo_definition}, the OP can be expressed~as \vspace{-0.2cm}
	\begin{align}
		P_{o}^{\text{id-w}}&(\gamma_{\mathrm{th}})=\Pr\left(A_{e2e}\leq \sqrt{\frac{\gamma_{\mathrm{th}}}{\gamma}}\right), 
	\end{align}
	or equivalently \vspace{-0.2cm}
	\begin{align}
		P_{o}^{\text{id-w}}&(\gamma_{\mathrm{th}})= F_{A_{e2e}}\left(\sqrt{\frac{\gamma_{\mathrm{th}}}{\gamma}}\right),
	\end{align}
	which with the aid of~\eqref{Eq:F_A_e2e_final_general} returns~\eqref{Eq:P_o_id_w}. This concludes the~proof.  
\end{IEEEproof}

For the special case in which disorientation and misalignment are experienced and $k_A\neq j \, m_A$ as well as $m_A\neq j \, k_A$ with $j=1,2, \cdots$, the OP can be evaluated as in~\eqref{Eq:P_o_s_id_w}, given at the top of the next~page.
\begin{figure*}
	\begin{align}
		P_{o,s}^{\text{id-w}}(\gamma_{\mathrm{th}})&=  \frac{\Xi^\zeta \Gamma\left(k_A-\frac{\zeta}{2}\right) \Gamma\left(m_A-\frac{\zeta}{2}\right)}{B_o^{\zeta} \Gamma\left(k_A\right) \Gamma\left(m_A\right)}
		+ \frac{\rho\, \Xi^{2k_A}}{k_A \zeta B_o^{2\zeta } }\frac{\Gamma\left(-k_A+m_A\right)}{\Gamma\left(k_A\right)\Gamma\left(m_A\right)}  \left(\frac{\gamma_{\mathrm{th}}}{\gamma}\right)^{k_A}\, \,_1F_2\left(k_A; 1+k_A, 1+k_A-m_A; \frac{\Xi^2}{B_o^2}\frac{\gamma_{\mathrm{th}}}{\gamma}\right)
		\nonumber \\ & 
		-  \frac{2 \Xi^{2k_A}}{ \left(2k_A-\rho\right) B_o^{2\zeta}} \frac{\Gamma\left(-k_A+m_A\right)}{\Gamma\left(k_A\right)\Gamma\left(m_A\right)}  \left(\frac{\gamma_{\mathrm{th}}}{\gamma}\right)^{k_A}\, \,_1F_2\left(k_A-\frac{\zeta}{2}; 1+k_A-m_A, 1+k_A-\frac{\zeta}{2}; \frac{\Xi^2}{B_o^2} \frac{\gamma_{\mathrm{th}}}{\gamma}\right)
		\nonumber \\ & 
		+\frac{2\rho  \Xi^{2m_A}}{m_A\zeta B_o^{2m_A}}\frac{\Gamma\left(k_A-m_A\right)}{\Gamma\left(k_A\right)\Gamma\left(m_A\right)}  \left(\frac{\gamma_{\mathrm{th}}}{\gamma}\right)^{m_A}\,\,_1F_2\left(m_A;1+m_A, 1-k_A+m_A; \frac{\Xi^2}{B_o^2}\frac{\gamma_{\mathrm{th}}}{\gamma}\right)
		\nonumber \\ &
		-\frac{2\rho\Xi^{2m_A}}{\left(2m_A-\zeta\right)\zeta B_o^{2m_A}}\frac{\Gamma\left(k_A-m_A\right)}{\Gamma\left(k_A\right)\Gamma\left(m_A\right)}  \left(\frac{\gamma_{\mathrm{th}}}{\gamma}\right)^{m_A} \, \,_1F_2\left(m_A - \frac{\zeta}{2}; 1-k_A+m_A, 1+m_A-\frac{\zeta}{2}; \frac{\Xi^2}{B_o^2} \frac{\gamma_{\mathrm{th}}}{\gamma}\right)
		\label{Eq:P_o_s_id_w} 
	\end{align}
\vspace{-0.3cm}
\hrulefill
\end{figure*}  
Moreover, the following lemma returns a high-SNR approximation for the OP.
\begin{lem}
	In presence of disorientation and misalignment, the OP can be approximated as in~\eqref{Eq:P_o_s_id_w_infty}, given at the top of the next page. 
	\begin{figure*}
	\begin{align}
		P_{o,s,\infty}^{\text{id-w}}(\gamma_{\mathrm{th}})&=  \frac{\Xi^\zeta \Gamma\left(k_A-\frac{\zeta}{2}\right) \Gamma\left(m_A-\frac{\zeta}{2}\right)}{B_o^{\zeta} \Gamma\left(k_A\right) \Gamma\left(m_A\right)}
		+ \frac{\rho\, \Xi^{2k_A}}{k_A \zeta B_o^{2\zeta } }\frac{\Gamma\left(-k_A+m_A\right)}{\Gamma\left(k_A\right)\Gamma\left(m_A\right)}  \left(\frac{\gamma_{\mathrm{th}}}{\gamma}\right)^{k_A}\, 
	%	\nonumber \\ & 
		-  \frac{2 \Xi^{2k_A}}{ \left(2k_A-\rho\right) B_o^{2\zeta}} \frac{\Gamma\left(-k_A+m_A\right)}{\Gamma\left(k_A\right)\Gamma\left(m_A\right)}  \left(\frac{\gamma_{\mathrm{th}}}{\gamma}\right)^{k_A}\, 
		\nonumber \\ & 
		+\frac{2\rho  \Xi^{2m_A}}{m_A\zeta B_o^{2m_A}}\frac{\Gamma\left(k_A-m_A\right)}{\Gamma\left(k_A\right)\Gamma\left(m_A\right)}  \left(\frac{\gamma_{\mathrm{th}}}{\gamma}\right)^{m_A}
		%\nonumber \\ &
		-\frac{2\rho\Xi^{2m_A}}{\left(2m_A-\zeta\right)\zeta B_o^{2m_A}}\frac{\Gamma\left(k_A-m_A\right)}{\Gamma\left(k_A\right)\Gamma\left(m_A\right)}  \left(\frac{\gamma_{\mathrm{th}}}{\gamma}\right)^{m_A} 
		\label{Eq:P_o_s_id_w_infty}
	\end{align}
\vspace{-0.2cm}
	\hrulefill 
	\end{figure*}
\end{lem}
\begin{IEEEproof}
	The proof of Lemma 5 follows the same steps as the one of Lemma 4. 
\end{IEEEproof}

The following corollary returns an OP floor.
\begin{cor}[OP floor] An OP floor exists that can be calculated~as \vspace{-0.2cm}
	\begin{align} 
		P_{o,f}^{\text{id-w}}(\gamma_{\mathrm{th}})&=  \frac{\Xi^\zeta \Gamma\left(k_A-\frac{\zeta}{2}\right) \Gamma\left(m_A-\frac{\zeta}{2}\right)}{B_o^{\zeta} \Gamma\left(k_A\right) \Gamma\left(m_A\right)}.
		\label{Eq:P_o_f}
	\end{align}
\end{cor}
\begin{IEEEproof}
	For $\gamma\to\infty$, it can be easily seen from~\eqref{Eq:P_o_s_id_w_infty} that $P_{o,s,\infty}^{\text{id-w}}(\gamma_{\mathrm{th}})\to P_{o,f}^{\text{id-w}}(\gamma_{\mathrm{th}})$. This concludes the proof.  
\end{IEEEproof}

\subsubsection{Diversity order}\label{SS:DO}
The following proposition returns the diversity order of RIS-assisted UAV wireless systems in the absence of transceivers RF front-end imperfections. 
\begin{prop} 
	The diversity order can be evaluated~as\vspace{-0.2cm}
	\begin{align}
		\mathcal{D}_{\text{id}}= \max\left({k_A}, {m_A}\right).
		\label{Eq:Diversity_order_ideal}
	\end{align}
\end{prop}
\begin{IEEEproof}
	From~\eqref{Eq:P_o_s_id_wo_approx} and~\eqref{Eq:P_o_s_id_w_infty}, it is evident that $\left(\frac{\gamma_{\mathrm{th}}}{\gamma}\right)^{k_A}$ and $\left(\frac{\gamma_{\mathrm{th}}}{\gamma}\right)^{m_A}$ respectively contribute with diversity order $k_A$ and $m_A$ in the asymptotic OP. As a result, the diversity order can be obtained as in~\eqref{Eq:Diversity_order_ideal}. This concludes the proof. 
\end{IEEEproof}

\vspace{-0.5cm}
\subsection{Non-ideal RF front-end}\label{SS:NIdeal_RF_FE} \vspace{-0.2cm}
This section focuses on extracting the outage performance analysis framework capable of assessing the effects of small-scale fading, disorientation and misalignment, and transceivers hardware imperfections. Specifically, Section~\ref{SS:SDNR} presents the instantaneous signal-to-distortion-plus-noise-ratio (SDNR), while, Section~\ref{SS:OP_WHI} derives closed-form formulas and high-SNR approximations. Finally, Section~\ref{SS:DO_HI} is devoted to the extraction of the diversity order.        

\subsubsection{Instantaneous SDNR}\label{SS:SDNR}  
For the case in which both the S and UAV transceivers suffer from RF front-end imperfection, the SDNR at the UAV can be expressed~as  \vspace{-0.2cm}
\begin{align}
	\gamma_u\hspace{-0.1cm} =\hspace{-0.1cm} \left\{ \hspace{-0.1cm}
	\begin{array}{c l}
		\frac{h_l^2 A^2 P_s}{\left(\kappa_s^2+\kappa_d^2\right) h_l^2 A^2 P_s + \sigma_w^2}, & \text{ without disorientation} \\  &  \text{ or misalignment}\\
		\frac{h_l^2 A_{e2e}^2 P_s}{\left(\kappa_s^2+\kappa_d^2\right) h_l^2 A_{e2e}^2 P_s + \sigma_w^2}, & \text{ with disorientation} \\  &  \text{ and misalignment}
	\end{array}\right.
\end{align}
or equivalently \vspace{-0.2cm}
\begin{align}
	\gamma_u\hspace{-0.1cm} =\hspace{-0.1cm} \left\{ 
	\begin{array}{c l}
		\frac{A^2}{\left(\kappa_s^2+\kappa_d^2\right) A^2 + \frac{1}{\gamma}}, & \text{ without disorientation} \\  &  \text{ or misalignment}\\
		\frac{A_{e2e}^2}{\left(\kappa_s^2+\kappa_d^2\right) A_{e2e}^2 + \frac{1}{\gamma}}, & \text{ with disorientation} \\  &  \text{ and misalignment}
	\end{array}\right..
	\label{Eq:gamma_u}
\end{align}

\subsubsection{OP}\label{SS:OP_WHI} 
In this section, closed-form expressions and high-SNR approximations for the OP for the case in which the S transmitter and the UAV receiver experience hardware imperfections are provided.

\underline{Without disorientation and misalignment}: The following proposition returns the OP in the absence of disorientation and misalignment.
\begin{prop}
	In the absence of both disorientation and misalignment, the OP can be evaluated~as \vspace{-0.2cm}
	\begin{align}
		P_o^{\text{wo}}\left(\gamma_{\mathrm{th}}\right) &\hspace{-0.12cm}=\hspace{-0.12cm} 
		\left\{ \hspace{-0.17cm}
		\begin{array}{c l}
			\frac{\mathrm{G}_{1, 3}^{2, 1}\left( \frac{\Xi^2}{1-\left(\kappa_s^2+\kappa_d^2\right)\gamma_{\mathrm{th}}} \frac{\gamma_{\mathrm{th}}}{\gamma}\left|\begin{array}{c} 1\\ k_A, m_A, 0 \end{array}\right.\right)}{\Gamma\left(k_A\right)\Gamma\left(m_A\right)}, & \\ 
			& \hspace{-2.9cm}\text{ for } \gamma_{\mathrm{th}}< \frac{1}{\kappa_s^2 + \kappa_d^2} \\
			1, & \hspace{-2.9cm} \text{ for } \gamma_{\mathrm{th}}\geq \frac{1}{\kappa_s^2 + \kappa_d^2}
		\end{array} 
		\right.
		\label{Eq:Po_wo_final}
	\end{align}
\end{prop}
\begin{IEEEproof}
	For brevity, the proof of Proposition 4 is provided in Appendix C. 
\end{IEEEproof}

\textbf{Remark 1.} From~\eqref{Eq:Po_wo_final}, it becomes evident that there exists a maximum SNR threshold that is equal to \vspace{-0.2cm}
\begin{align}
	\gamma_{\mathrm{th}}^{m}=\frac{1}{\kappa_s^2 + \kappa_d^2},
	\label{Eq:gamma_th_m}
\end{align}
beyond which the OP becomes equal to $1$. Interestingly,~$\gamma_{\mathrm{th}}^{m}$ only depends on the levels of hardware imperfections of the S's transmitter and UAV's receiver.  

In the special case in which $k_A\neq j m_A$ and $m_A \neq j k_A$, with $j=1, 2, 3, \cdots$, by following the same process as in the proof of Proposition 4, the OP can be alternatively evaluated as in~\eqref{Eq:P_o_s_wo}, given at the top of the next~page. 
\begin{figure*}
	\begin{align}
	P_{o,s}^{\text{wo}}(\gamma_{\mathrm{th}}) \hspace{-0.1cm} = \hspace{-0.1cm}\left\{ \hspace{-0.2cm}
	\begin{array}{c l}
	\frac{\pi \csc\left(\left(k_A-m_A\right)\pi\right) \Xi^{2m_A} }{\Gamma\left(k_A\right)  \Gamma\left(m_A+1\right)\Gamma\left(1-k_A+m_A\right)}
	\hspace{-0.1cm}\left(\frac{1}{1-\left(\kappa_s^2+\kappa_d^2\right)\gamma_{\mathrm{th}}}\right)^{m_A} 
	\hspace{-0.1cm} \left(\frac{\gamma_{\mathrm{th}}}{\gamma}\right)^{m_A}
	% & \\
	 \hspace{-0.1cm}\,_1F_2\left(m_A;1+m_A, 1-k_A+m_A; \Xi^2 \frac{1}{1-\left(\kappa_s^2+\kappa_d^2\right)\gamma_{\mathrm{th}}} \frac{\gamma_{\mathrm{th}}}{\gamma}\right)
	 & \\   
	- \frac{\pi \csc\left(\left(k_A-m_A\right)\pi\right) \Xi^{2k_A} }{\Gamma\left(m_A\right)  \Gamma\left(k_A+1\right)\Gamma\left(1+k_A-m_A\right)} \left(\frac{1}{1-\left(\kappa_s^2+\kappa_d^2\right)\gamma_{\mathrm{th}}}\right)^{k_A}\left(\frac{\gamma_{\mathrm{th}}}{\gamma}\right)^{k_A}
	%& \\
	\,_1F_2\left(k_A;1+k_A, 1+k_A-m_A; \Xi^2 \frac{1}{1-\left(\kappa_s^2+\kappa_d^2\right)\gamma_{\mathrm{th}}} \frac{\gamma_{\mathrm{th}}}{\gamma}\right)
	& \\ & \hspace{-5cm}\text{ for } \gamma_{\mathrm{th}}<\frac{1}{\kappa_{s}^2 + \kappa_d^2} \\
	1, & \hspace{-5cm}\text{ for } \gamma_{\mathrm{th}}\geq\frac{1}{\kappa_{s}^2 + \kappa_d^2}
	\end{array}
	\right.
	\label{Eq:P_o_s_wo}
	\end{align}
 \vspace{-0.5cm}
	\hrulefill
\end{figure*} 
For this case, the following lemma returns a high-SNR approximation. 
\begin{lem}
	In the high-SNR regime, the OP can be approximated as in~\eqref{Eq:P_o_s_wo_infty}, given at the top of the next page. 
\begin{figure*}
	\begin{align}
		P_{o,\infty}^{\text{wo}}(\gamma_{\mathrm{th}}) \hspace{-0.1cm}= \hspace{-0.1cm}\left\{ \hspace{-0.1cm}
		\begin{array}{c l}
			\hspace{-0.1cm}\frac{\pi \csc\left(\left(k_A-m_A\right)\pi\right) \Xi^{2m_A} }{\Gamma\left(k_A\right)  \Gamma\left(m_A+1\right)\Gamma\left(1-k_A+m_A\right)}
			\left(\frac{1}{1-\left(\kappa_s^2+\kappa_d^2\right)\gamma_{\mathrm{th}}}\right)^{m_A} 
			\left(\frac{\gamma_{\mathrm{th}}}{\gamma}\right)^{m_A}
			% & \\
			%& \\   
			\hspace{-0.2cm}- \frac{\pi \csc\left(\left(k_A-m_A\right)\pi\right) \Xi^{2k_A} }{\Gamma\left(m_A\right)  \Gamma\left(k_A+1\right)\Gamma\left(1+k_A-m_A\right)} \hspace{-0.1cm}\left(\frac{1}{1-\left(\kappa_s^2+\kappa_d^2\right)\gamma_{\mathrm{th}}}\right)^{k_A}\hspace{-0.1cm}\left(\frac{\gamma_{\mathrm{th}}}{\gamma}\right)^{k_A}
			%& \\
			& \\ & \hspace{-5cm}\text{ for } \gamma_{\mathrm{th}}<\frac{1}{\kappa_{s}^2 + \kappa_d^2} \\
			1, & \hspace{-5cm}\text{ for } \gamma_{\mathrm{th}}\geq\frac{1}{\kappa_{s}^2 + \kappa_d^2}
		\end{array}
		\right.
		\label{Eq:P_o_s_wo_infty}
	\end{align}
\vspace{-0.5cm}
	\hrulefill
\end{figure*}
\end{lem}
\begin{IEEEproof}
	The proof of Lemma 6 follows the same steps as the proof of Lemma 4. 
\end{IEEEproof}

\underline{With disorientation and misalignment}: The following proposition returns the OP in the presence of disorientation and misalignment.
\begin{prop}
	In the presence of disorientation and misalignment, the OP can be calculated~as in~\eqref{Eq:P_o_w}, given at the top of the next page. 
	\begin{figure*}
	\begin{align}
		P_{o}^{\text{w}}&(\gamma_{\mathrm{th}})=
		\left\{
		\begin{array}{c l}
		 \frac{\zeta \mathrm{G}_{5,3}^{1,4}\left(\left.\begin{array}{c}1-k_A, 1-m_A, \frac{1-\zeta}{2}, \frac{2-\zeta}{2},1 \\ 0, \frac{1-\zeta}{2}, - \frac{\zeta}{2}\end{array}\right| \frac{B_o^2}{\Xi^2} \frac{\gamma}{\gamma_{\mathrm{th}}}\right)}{2\Gamma\left(k_A\right)\Gamma\left(m_A\right)}, & \text{ for } \gamma_{\mathrm{th}}< \frac{1}{\kappa_s^2 + \kappa_d^2} \\
		 1, & \text{ for } \gamma_{\mathrm{th}} \geq \frac{1}{\kappa_s^2 + \kappa_d^2}
		\end{array}
		\right.
		\label{Eq:P_o_w}
	\end{align}
\vspace{-0.5cm}
	\hrulefill
	\end{figure*}
\end{prop} 
\begin{IEEEproof}
	For brevity, the proof of proposition 5 is given in Appendix D.  
\end{IEEEproof}

In the special case in which $k_A\neq j\, m_A$ and $m_A \neq j\, k_A$ with $j=1,2, \cdots$, by following the same steps as in the proof of Proposition 5 and employing~\eqref{Eq:F_A_2e2_final_special}, the OP can be equivalently written as in~\eqref{Eq:P_o_s_w}, given at the top of the next~page. 
\begin{figure*}
	\begin{align}
		P_{o,s}^{\text{w}}(\gamma_{\mathrm{th}}) = \left\{
		\begin{array}{l l}
			\frac{\Xi^\zeta \Gamma\left(k_A-\frac{\zeta}{2}\right) \Gamma\left(m_A-\frac{\zeta}{2}\right)}{B_o^{\zeta} \Gamma\left(k_A\right) \Gamma\left(m_A\right)}
			& \\ 
			+ \frac{\rho\, \Xi^{2k_A}}{k_A \zeta B_o^{2\zeta } }\frac{\Gamma\left(-k_A+m_A\right)}{\Gamma\left(k_A\right)\Gamma\left(m_A\right)} 
			\left(\frac{1}{1-\left(\kappa_s^2+\kappa_d^2\right)\gamma_{\mathrm{th}}}\right)^{k_A} 
			\left(\frac{\gamma_{\mathrm{th}}}{\gamma}\right)^{k_A}\, \,_1F_2\left(k_A; 1+k_A, 1+k_A-m_A; \frac{\Xi^2}{B_o^2} \frac{1}{1-\left(\kappa_s^2+\kappa_d^2\right)\gamma_{\mathrm{th}}} \frac{\gamma_{\mathrm{th}}}{\gamma}\right)
				& \\ 
			-  \frac{2 \Xi^{2k_A}}{ \left(2k_A-\rho\right) B_o^{2\zeta}} \frac{\Gamma\left(-k_A+m_A\right)}{\Gamma\left(k_A\right)\Gamma\left(m_A\right)}  \left(\frac{1}{1-\left(\kappa_s^2+\kappa_d^2\right)\gamma_{\mathrm{th}}}\right)^{k_A} 
			\left(\frac{\gamma_{\mathrm{th}}}{\gamma}\right)^{k_A}\, \,_1F_2\left(k_A-\frac{\zeta}{2}; 1+k_A-m_A, 1+k_A-\frac{\zeta}{2}; \frac{\Xi^2}{B_o^2} \frac{1}{1-\left(\kappa_s^2+\kappa_d^2\right)\gamma_{\mathrm{th}}} \frac{\gamma_{\mathrm{th}}}{\gamma}\right)
				& \\ 
			+\frac{2\rho  \Xi^{2m_A}}{m_A\zeta B_o^{2m_A}}\frac{\Gamma\left(k_A-m_A\right)}{\Gamma\left(k_A\right)\Gamma\left(m_A\right)}
			\left(\frac{1}{1-\left(\kappa_s^2+\kappa_d^2\right)\gamma_{\mathrm{th}}}\right)^{m_A}   \left(\frac{\gamma_{\mathrm{th}}}{\gamma}\right)^{m_A}\,\,_1F_2\left(m_A;1+m_A, 1-k_A+m_A; \frac{\Xi^2}{B_o^2}\frac{1}{1-\left(\kappa_s^2+\kappa_d^2\right)\gamma_{\mathrm{th}}} \frac{\gamma_{\mathrm{th}}}{\gamma}\right)
				& \\ 
			-\frac{2\rho\Xi^{2m_A}}{\left(2m_A-\zeta\right)\zeta B_o^{2m_A}}\frac{\Gamma\left(k_A-m_A\right)}{\Gamma\left(k_A\right)\Gamma\left(m_A\right)} \left(\frac{1}{1-\left(\kappa_s^2+\kappa_d^2\right)\gamma_{\mathrm{th}}}\right)^{m_A}  \left(\frac{\gamma_{\mathrm{th}}}{\gamma}\right)^{m_A} & \\ \hspace{+2cm}\times \,_1F_2\left(m_A - \frac{\zeta}{2}; 1-k_A+m_A, 1+m_A-\frac{\zeta}{2}; \frac{\Xi^2}{B_o^2} \frac{1}{1-\left(\kappa_s^2+\kappa_d^2\right)\gamma_{\mathrm{th}}} \frac{\gamma_{\mathrm{th}}}{\gamma}\right),
			 & \hspace{-5cm}\text{ for } \gamma_{\mathrm{th}}<\frac{1}{\kappa_{s}^2 + \kappa_d^2} \\
			1, & \hspace{-5cm}\text{ for } \gamma_{\mathrm{th}}\geq\frac{1}{\kappa_{s}^2 + \kappa_d^2}
		\end{array}
		\right.
		\label{Eq:P_o_s_w}
	\end{align}
\vspace{-0.5cm}
	\hrulefill
\end{figure*} 
Likewise, in this case, the following lemma returns a high-SNR approximation. 
\begin{lem}
	In the high SNR regime, the OP can be approximated as in~\eqref{Eq:P_o_s_w_approx}, given at the top of the next page. 
\end{lem} 
\begin{IEEEproof}
	Lemma 7 can be proven by following the same steps as Lemma 4. 
\end{IEEEproof}
\begin{figure*}
	\begin{align}
		P_{o,\infty}^{\text{w}}(\gamma_{\mathrm{th}}) = \left\{
		\begin{array}{l l}
			\frac{\Xi^\zeta \Gamma\left(k_A-\frac{\zeta}{2}\right) \Gamma\left(m_A-\frac{\zeta}{2}\right)}{B_o^{\zeta} \Gamma\left(k_A\right) \Gamma\left(m_A\right)}
			+ \frac{\rho\, \Xi^{2k_A}}{k_A \zeta B_o^{2\zeta } }\frac{\Gamma\left(-k_A+m_A\right)}{\Gamma\left(k_A\right)\Gamma\left(m_A\right)} 
			\left(\frac{1}{1-\left(\kappa_s^2+\kappa_d^2\right)\gamma_{\mathrm{th}}}\right)^{k_A} 
			\left(\frac{\gamma_{\mathrm{th}}}{\gamma}\right)^{k_A} 
			& \\ 
			-  \frac{2 \Xi^{2k_A}}{ \left(2k_A-\rho\right) B_o^{2\zeta}} \frac{\Gamma\left(-k_A+m_A\right)}{\Gamma\left(k_A\right)\Gamma\left(m_A\right)}  \left(\frac{1}{1-\left(\kappa_s^2+\kappa_d^2\right)\gamma_{\mathrm{th}}}\right)^{k_A} 
			\left(\frac{\gamma_{\mathrm{th}}}{\gamma}\right)^{k_A}
			+\frac{2\rho  \Xi^{2m_A}}{m_A\zeta B_o^{2m_A}}\frac{\Gamma\left(k_A-m_A\right)}{\Gamma\left(k_A\right)\Gamma\left(m_A\right)}
			\left(\frac{1}{1-\left(\kappa_s^2+\kappa_d^2\right)\gamma_{\mathrm{th}}}\right)^{m_A}   \left(\frac{\gamma_{\mathrm{th}}}{\gamma}\right)^{m_A}
			& \\ 
			-\frac{2\rho\Xi^{2m_A}}{\left(2m_A-\zeta\right)\zeta B_o^{2m_A}}\frac{\Gamma\left(k_A-m_A\right)}{\Gamma\left(k_A\right)\Gamma\left(m_A\right)} \left(\frac{1}{1-\left(\kappa_s^2+\kappa_d^2\right)\gamma_{\mathrm{th}}}\right)^{m_A}  \left(\frac{\gamma_{\mathrm{th}}}{\gamma}\right)^{m_A}, 
			& \hspace{-5cm}\text{ for } \gamma_{\mathrm{th}}<\frac{1}{\kappa_{s}^2 + \kappa_d^2} \\
			1, & \hspace{-5cm}\text{ for } \gamma_{\mathrm{th}}\geq\frac{1}{\kappa_{s}^2 + \kappa_d^2}
		\end{array}
		\right.
		\label{Eq:P_o_s_w_approx}
	\end{align}
	\vspace{-0.5cm}
	\hrulefill
\end{figure*} 

The following corollary returns the OP floor.
\begin{cor}[OP floor] The OP probability presents a floor that can be evaluated~as \vspace{-0.2cm}
	\begin{align}
		P_{o,f}^{w} = \left\{ \begin{array}{l l} \frac{\Xi^\zeta \Gamma\left(k_A-\frac{\zeta}{2}\right) \Gamma\left(m_A-\frac{\zeta}{2}\right)}{B_o^{\zeta} \Gamma\left(k_A\right) \Gamma\left(m_A\right)}, & \text{ for } \gamma_{\mathrm{th}}<\frac{1}{\kappa_{s}^2 + \kappa_d^2} \\
			1,  & \text{ for } \gamma_{\mathrm{th}}\geq\frac{1}{\kappa_{s}^2 + \kappa_d^2} \end{array}\right.. 
		\label{Eq:P_of_w}
	\end{align}
\end{cor}
\begin{IEEEproof}
	For $\gamma\to\infty$, $\left(\frac{\gamma_{\mathrm{th}}}{\gamma}\right)\to 0$; thus, \vspace{-0.2cm}
	\begin{align}
	\lim_{\gamma\to\infty} P_{o,\infty}^{\text{w}}(\gamma_{\mathrm{th}}) \hspace{-0.1cm}=\hspace{-0.1cm} \left\{ \hspace{-0.1cm} \begin{array}{l l} \frac{\Xi^\zeta \Gamma\left(k_A-\frac{\zeta}{2}\right) \Gamma\left(m_A-\frac{\zeta}{2}\right)}{B_o^{\zeta} \Gamma\left(k_A\right) \Gamma\left(m_A\right)}, & \text{ for } \gamma_{\mathrm{th}}<\frac{1}{\kappa_{s}^2 + \kappa_d^2} \\
		1,  & \text{ for } \gamma_{\mathrm{th}}\geq\frac{1}{\kappa_{s}^2 + \kappa_d^2} \end{array}\right.. 
	\end{align}
	This concludes the proof. 
\end{IEEEproof}

\textbf{Remark 2.} For $\gamma_{\mathrm{th}}<\frac{1}{\kappa_{s}^2 + \kappa_d^2}$, by comparing~\eqref{Eq:P_o_f} and~\eqref{Eq:P_of_w}, it becomes apparent that the OP floor depends on the fading conditions, the RIS characteristics, which are captured by $k_A$ and $m_A$, as well as the level of disorientation and~misalignment. Moreover, it is independent of the level of transceivers hardware imperfections.  

\subsubsection{Diversity order}\label{SS:DO_HI} 
The following proposition returns the diversity order of RIS-assisted UAV wireless systems in the presence of transceivers RF front-end imperfections.
\begin{prop}
	The diversity order can be calculated~as \vspace{-0.2cm}
	\begin{align}
		\mathcal{D} = \max\left(k_A, m_A\right). 
		\label{Eq:Diversity_order_non_ideal}
	\end{align}
\end{prop}
\begin{IEEEproof}
From~\eqref{Eq:P_o_s_wo_infty} and~\eqref{Eq:P_o_s_w_approx}, it is evident that $\left(\frac{1}{\gamma}\right)^{k_A}$ and $\left(\frac{1}{\gamma}\right)^{m_A}$ respectively contribute with diversity order $k_A$ and $m_A$ in the asymptotic OP. As a result, the diversity order can be obtained as in~\eqref{Eq:Diversity_order_non_ideal}. This concludes the proof.
\end{IEEEproof}

\textbf{Remark 3.} By comparing~\eqref{Eq:Diversity_order_ideal} and~\eqref{Eq:Diversity_order_non_ideal}, it becomes evident that the transceivers hardware imperfections do not affect the diversity order of RIS-assisted UAV wireless systems.

\section{Results \& Discussion}\label{S:Results} \vspace{-0.2cm}

This section is devoted to verify the theoretical framework that has been presented in Section~\ref{S:PA}, by means of Monte Carlo simulations and provide an insightful discussion of the effects of fading, disorientation and misalignment in RIS-assisted UAV wireless systems. Along these lines, the following insightful scenario is considered. It is assumed that S-RIS channel coefficients follow independent and identical Nakagami-$m$ distributions with spread parameters $\Omega=1$; thus, $M=1$, $a_{1}^{(1)}=\frac{m^m}{\Gamma(m)}$, $b_1^{(1)}=m$, and $c_1=m$, where $m$ represents the shape parameter. The Rice distribution is used to model the channel coefficients  of the RIS-UAV link. For an accurate approximation of the Rice distribution, we select $K=20$, \vspace{-0.3cm}
\begin{align}
	a_{k}^{(2)} = \frac{\delta\left(K_r, k\right)}{\sum_{k_1=1}^K \delta\left(K_r, k_1\right) \Gamma\left(b_{k_1}^{(2)}\right) c_2^{-b_{k_1}^{(2)}}},
	\label{Eq:a_k_rice}
\end{align}    
where \vspace{-0.3cm}
\begin{align}
	\delta\left(K_r, k\right) = \frac{K_r^{k-1}\left(1+K_r\right)^k}{\exp\left(K_1\right) \left( (n-1)!\right)^2},
\end{align}
while $c_2=1+K_r$, and $b_{k}^{(2)}=k$. Moreover, note that in~\eqref{Eq:a_k_rice}, $K_r$ stands for the $K-$factor of the Rice distribution.

The rest of this section is structured as follows: In Section~\ref{SS:Ideal_RF_front_end_results}, numerical results that assess the joint impact of fading as well as the disorientation and misalignment in RIS-assisted UAV wireless systems, which do not experience transceiver hardware imperfections, are provided, while, Section~\ref{Eq:Non_ideal_RF_FE_results} is focused on presenting the corresponding results that quantify the joint impact of fading, disorientation and  misalignment as well as transceiver hardware imperfections. Of note, in what follows, continuous lines are used to denote results that obtained by the analytical expressions, while markers are used for simulations.      

\vspace{-0.2cm}
\subsection{Ideal RF front-end}\label{SS:Ideal_RF_front_end_results}
\subsubsection{Without disorientation and misalignment} \label{SS:wo_dis_orientation}
In section, we focus on the quantification of the effect of small-scale fading. In this direction, the joint effects of disorientation and misalignment are neglected. Likewise, both the S transmitter and UAV receiver are considered to be equipped with ideal RF front-end chains. 

\begin{figure}
	\centering\includegraphics[width=0.5\linewidth,trim=0 0 0 0,clip=false]{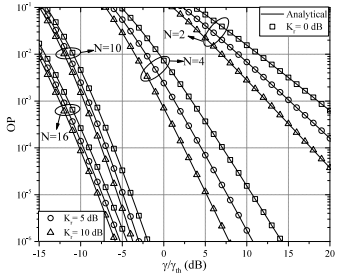}
	\vspace{-0.2cm}
	\caption{OP vs $\gamma_{s}/\gamma_{\mathrm{th}}$, for different values of $K_r$ and $N$.}
	\label{Fig:OP_vs_g_N_K}
	\vspace{-0.2cm}
\end{figure}

Figure~\ref{Fig:OP_vs_g_N_K} illustrates the OP as a function of $\gamma_{s}/\gamma_{\mathrm{th}}$, for different values of $K_r$ and $N$, assuming $m=1$. Analytical and simulations results coincide verifying the theoretical framework. Moreover, as expected, for given $K_r$ and $N$, as $\gamma_{s}/\gamma_{\mathrm{th}}$ increases the OP decreases. For example, for $N=4$ and $K_r=5\text{ }\mathrm{dB}$, the outage probability decreases for more than one order of magnitude, as $\gamma_{s}/\gamma_{\mathrm{th}}$ increases from $1$ to $5\text{ }\mathrm{dB}$. Likewise, for given $\gamma_{s}/\gamma_{\mathrm{th}}$ and $N$, as $K_r$ increases, i.e. as the RIS-D direct link becomes stronger, the outage performance improves. For instance, for $N=4$ and $\gamma_{s}/\gamma_{\mathrm{th}}=0\text{ }\mathrm{dB}$, the OP decreases for about one order of magnitude, as $K_r$ increases from $0$ to $10\text{ }\mathrm{dB}$. This indicates the importance of accurately modeling fading, when assessing the performance of RIS-assisted UAV wireless systems. Finally, for fixed $\gamma_{s}/\gamma_{\mathrm{th}}$ and $K_r$, as $N$ increases, the diversity order increases; hence, the OP decreases.  

\begin{figure}
	\centering\includegraphics[width=0.5\linewidth,trim=0 0 0 0,clip=false]{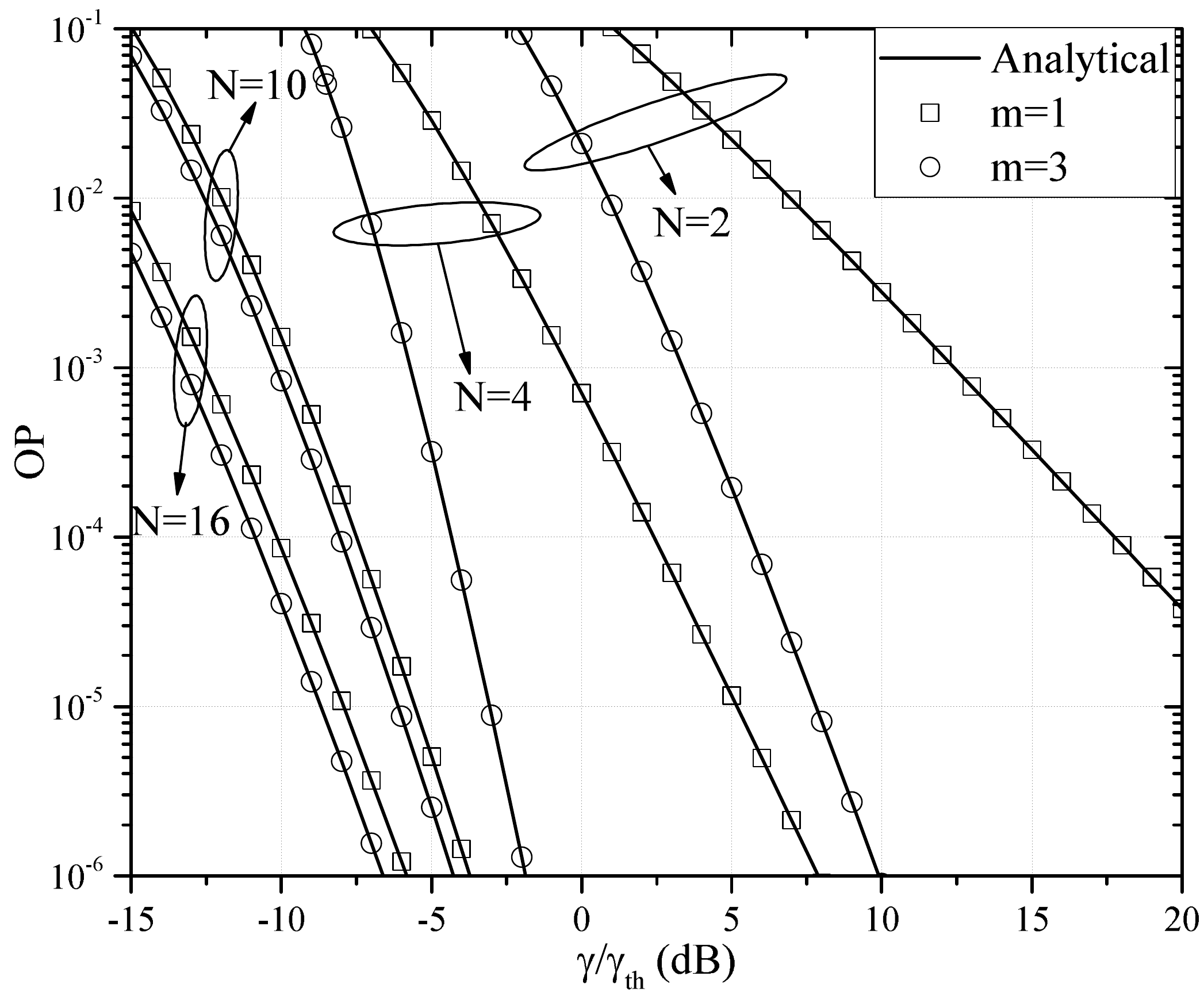}
	\vspace{-0.2cm}
	\caption{OP vs $\gamma_{s}/\gamma_{\mathrm{th}}$, for different values of $N$ and $m$.}
	\label{Fig:OP_vs_g_N_m}
	\vspace{-0.2cm}
\end{figure}

Figure~\ref{Fig:OP_vs_g_N_m} depicts the OP as a function of $\gamma_{s}/\gamma_{\mathrm{th}}$, for different values of $K_r$ and $m$, assuming $K_r=10\text{ }\mathrm{dB}$. Again, for a given $N$ and $m$, as $\gamma_{s}/\gamma_{\mathrm{th}}$, the OP decreases. For example, for $N=4$ and $m=3$, the outage probability decreases for about $10$ times, as $\gamma_{s}/\gamma_{\mathrm{th}}$ increases from $-7$ to $-5\text{ }\mathrm{dB}$. Likewise, for given $N$ and $\gamma_{s}/\gamma_{\mathrm{th}}$, as $m$ increases, the strength of the S-RIS direct link increases; thus, the outage performance improves. For instance, for $N=4$ and $\gamma_{s}/\gamma_{\mathrm{th}}=-2\text{ }\mathrm{dB}$, the OP decreases for more than three orders of magnitude, as $m$ increases from $1$ to $3$. From this example, the importance of accurately modeling the S-RIS link is revealed. Finally, for given $m$ and $\gamma_{s}/\gamma_{\mathrm{th}}$, as $N$ increases, the diversity order also increases; thus, the OP decreases.    

%\begin{figure}
%	\centering\includegraphics[width=0.5\linewidth,trim=0 0 0 0,clip=false]{images/Graph34.eps}
%	\vspace{-0.2cm}
%	\caption{OP vs $m_1$ and $m_2$, assuming $N=16$ and $\gamma_{s}/\gamma_{\mathrm{th}}=-10\text{ }\mathrm{dB}$.}
%	\label{Fig:OP_vs_m1_m2}
%\end{figure}      

%Figure~\ref{Fig:OP_vs_m1_m2} plots the OP as a function of $m_1$ and $m_2$, assuming that all the S-RIS channels can be modeled by independent and identical Nakagami-$m$ distributions with shape parameter $m_1$. Similarly, the RIS-D channels are models as  independent and identical Nakagami-$m$ distributed random variables with shape parameter $m_2$. Moreover, $N=16$ and $\gamma_{s}/\gamma_{\mathrm{th}}$ is set to $-10\text{ }\mathrm{dB}$. As expected, for a given $m_1$, as $m_2$ increases, the outage performance improves. Likewise, for a fixed $m_2$, as $m_1$ increases, the OP decreases. Finally, from this figure, it becomes evident that there exists a performance symmetry towards $m_1=m_2$. In other words, a system with $m_1=\alpha_1$ and $m_2=\alpha_2$ achieves the same performance as the one with $m_1=\alpha_2$ and $m_2=\alpha_2$, under the same $\gamma_{s}/\gamma_{\mathrm{th}}$.         

\subsubsection{With disorientation and misalignment} \label{SS:with_dis_orientation}

This section focuses on quantifying the joint impact of disorientation, misalignment and fading on the performance of RIS-assisted UAV wireless systems. The following  scenario is considered. Unless otherwise stated, the transmission distance between the RIS and the UAV is set to $L_2=5\text{ }\mathrm{m}$, whereas the beam-waist at the UAV reception plane is $w_o=1\text{ }\mathrm{mm}$. The transmission frequency is set to $f=100\text{ }\mathrm{GHz}$, and the index of the reflection structure parameter is  $C_n^2=2.3\times 10^{-9}\text{ }\mathrm{m}^{2/3}$. Finally, $\theta=\frac{7\pi}{4}$, $\phi=\frac{2\pi}{3}$, and $\alpha=10\text{ }\mathrm{cm}$.

\begin{figure}
	\centering\includegraphics[width=0.5\linewidth,trim=0 0 0 0,clip=false]{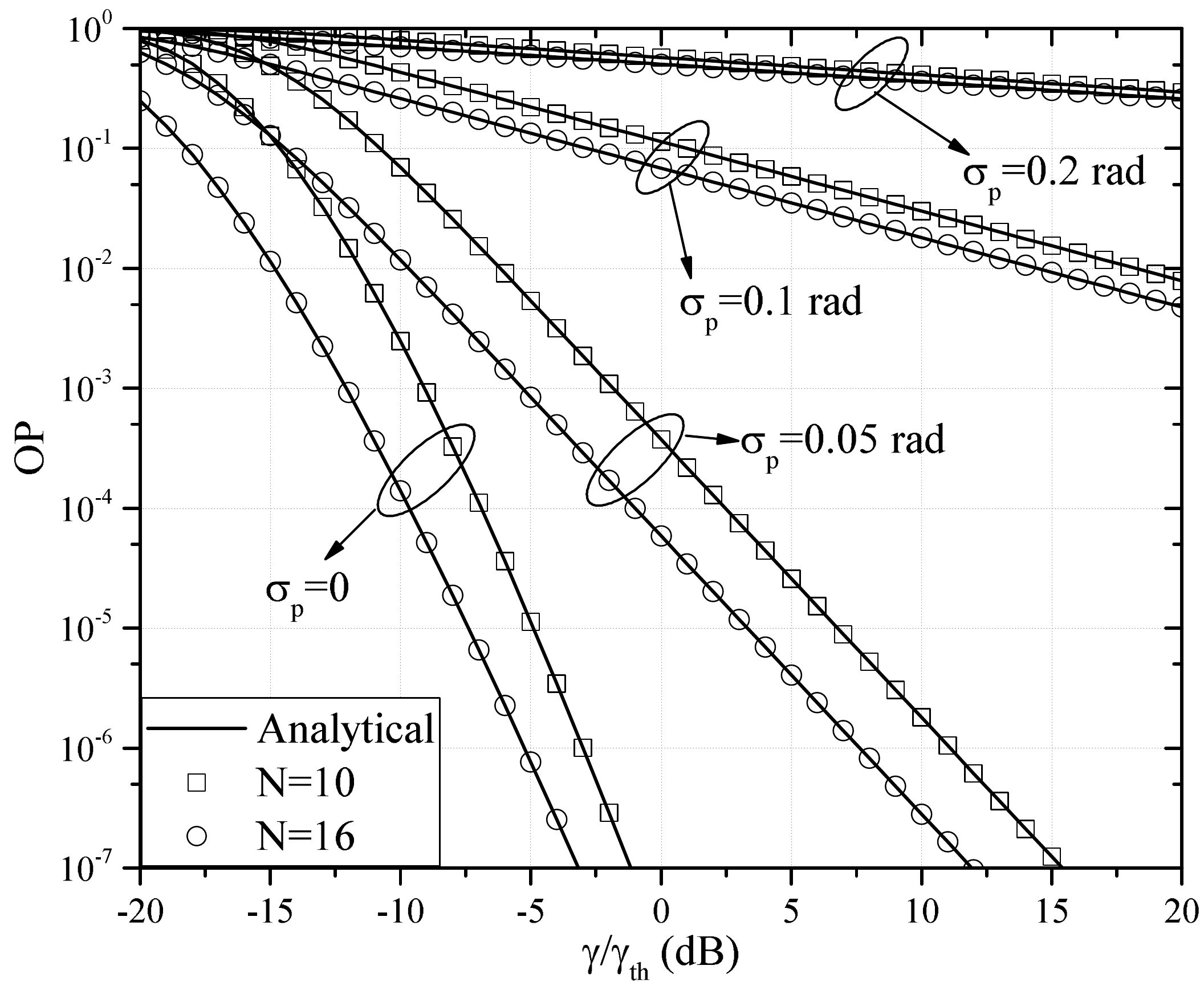}
	\vspace{-0.2cm}
	\caption{OP vs $\gamma/\gamma_{th}$ for different values of $\sigma_p$ and $N$.}
	\label{Fig:OP_vs_g_gth_N_sp}
	\vspace{-0.2cm}
\end{figure}  

Figure~\ref{Fig:OP_vs_g_gth_N_sp} depicts the OP as a function of $\gamma/\gamma_{th}$ for different values of $\sigma_p$ and $N$, assuming $\sigma_0=d_x=0$, $K_{r}=5\text{ }\mathrm{dB}$, and $m=1$. As a benchmark, the case in which the RIS-UAV link suffers from neither misalignment nor disorientation, i.e. $\sigma_p=0$, is illustrated. From this figure, we observe that, for fixed $\sigma_p$ and $N$, as $\gamma/\gamma_{\mathrm{th}}$ increases, the outage performance improves. Moreover, for given $\sigma_p$ and $\gamma/\gamma_{\mathrm{th}}$, as $N$ increases, according to~\eqref{Eq:Diversity_order_ideal}, the diversity order also  increases; hence, the OP decreases. Finally, for fixed $\gamma/\gamma_{\mathrm{th}}$ and $N$, as $\sigma_p$ increases, the OP also increases. For example, for $\gamma/\gamma_{\mathrm{th}}=-5\text{ }\mathrm{dB}$ and $N=10$, the OP increases for more than $2.5$ orders of magnitude, as $\sigma_p$ increases from $0$ to $0.05\text{ }\mathrm{rad}$. This example reveals the importance of accounting the impact of disorientation and misalignment when assessing the performance of RIS-assisted UAV wireless systems.

\begin{figure}
	\centering\includegraphics[width=0.5\linewidth,trim=0 0 0 0,clip=false]{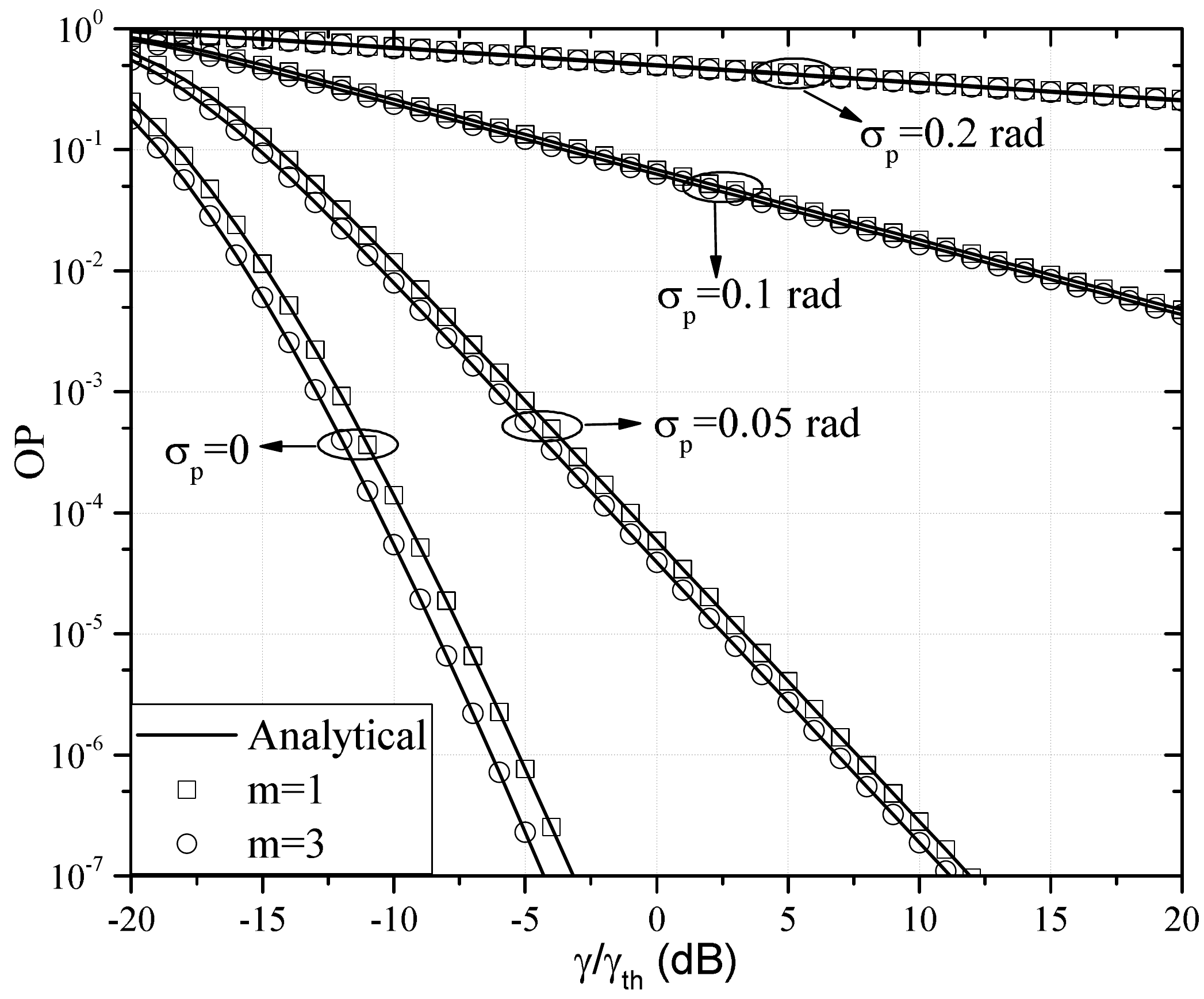}
	\vspace{-0.2cm}
	\caption{OP vs $\gamma/\gamma_{th}$ for different values of $\sigma_p$ and $m$.}
	\label{Fig:OP_vs_g_gth_m_sp}
	\vspace{-0.2cm}
\end{figure}   

Figure~\ref{Fig:OP_vs_g_gth_m_sp} illustrates the joint impact of S-RIS fading, disorientation and misalignment in the outage performance of the RIS-assisted UAV wireless systems. In more detail, the OP is plotted against the  $\gamma/\gamma_{th}$ for different values of $\sigma_p$ and $m$, assuming $\sigma_0=d_x=0$, $K_{r}=5\text{ }\mathrm{dB}$, and $N=16$. As expected, for given $\gamma/\gamma_{th}$ and $\sigma_p$, as $m$ increases, the OP decreases. For instance, for $\gamma/\gamma_{th}=5\text{ }\mathrm{dB}$ and $\sigma_p=0.05\text{ }\mathrm{rad}$, the OP decreases from $4.07\times 10^{-6}$ to $2.71\times 10^{-6}$, as $m$ increases from $1$ to $5$. Finally, from this figure, the detrimental effect of disorientation and misalignment becomes evident. In particular, as $\sigma_p$ increases the OP significantly increases. 

\begin{figure}
	\centering\includegraphics[width=0.5\linewidth,trim=0 0 0 0,clip=false]{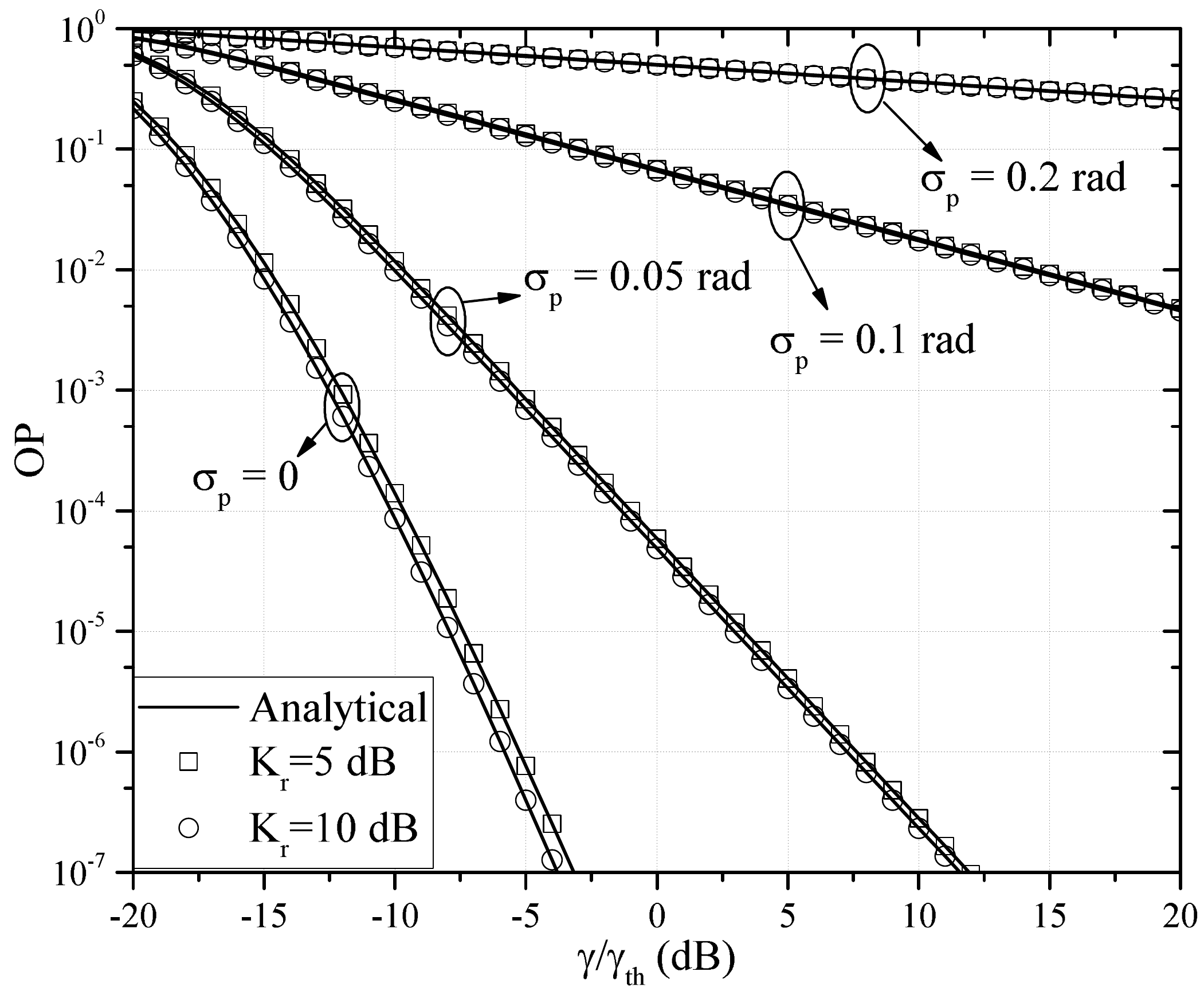}
	\vspace{-0.2cm}
	\caption{OP vs $\gamma/\gamma_{th}$ for different values of $\sigma_p$ and $K_r$.}
	\label{Fig:OP_vs_g_gth_Kr_sp}
	\vspace{-0.2cm}
\end{figure} 

Figure~\ref{Fig:OP_vs_g_gth_Kr_sp} demonstrates the joint impact of RIS-UAV fading, disorientation and misalignment in the outage performance of the RIS-assisted UAV wireless systems. Specifically, the OP is given as a function of OP vs $\gamma/\gamma_{th}$ for different values of $\sigma_p$ and $K_r$, assuming $\sigma_0=d_x=0$, $m=1$, and $N=16$. Again, in this figure analytical results agree with the simulations; thus, the theoretical framework is verified. Likewise, it becomes apparent from this figure that in the high-$\sigma_p$ regime, the phenomenon that dominates the performance of RIS-assisted UAV wireless systems is disorientation and misalignment and not the strength of the line-of-sight component. In more detail, we observe that, for a given $\gamma/\gamma_{th}$, in the high-$\sigma_p$ regime, no significant improvement of the outage performance occurs as $K_r$ increases. This indicates the importance of accurately modeling misalignment and disorientation.

\begin{figure}
	\begin{minipage}{0.48\linewidth}
	\centering\includegraphics[width=1\linewidth,trim=0 0 0 0,clip=false]{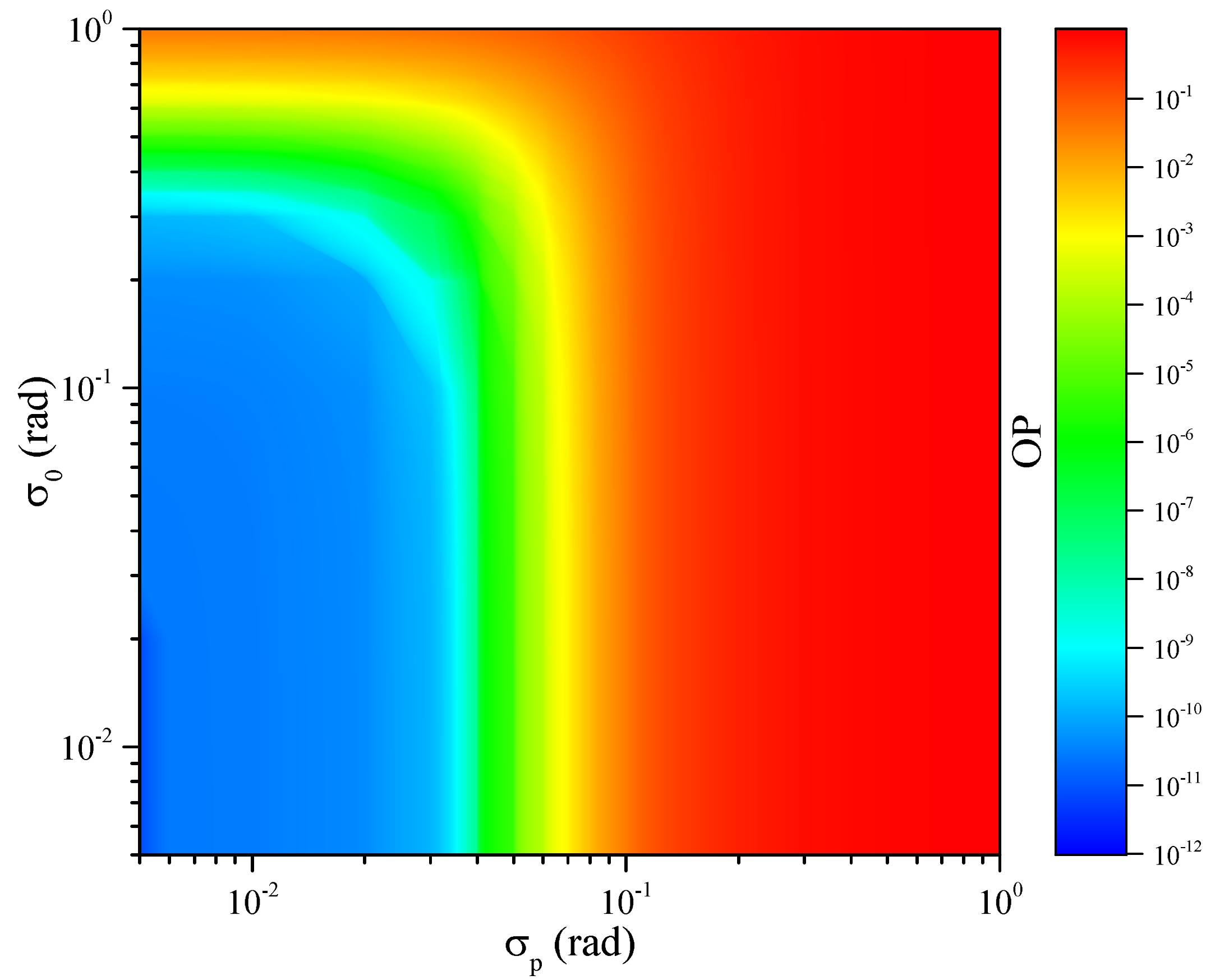}\\
	\vspace{-0.2cm}
	(a)
	\end{minipage}
	\begin{minipage}{0.48\linewidth}
	\includegraphics[width=1\linewidth,trim=0 0 0 0,clip=false]{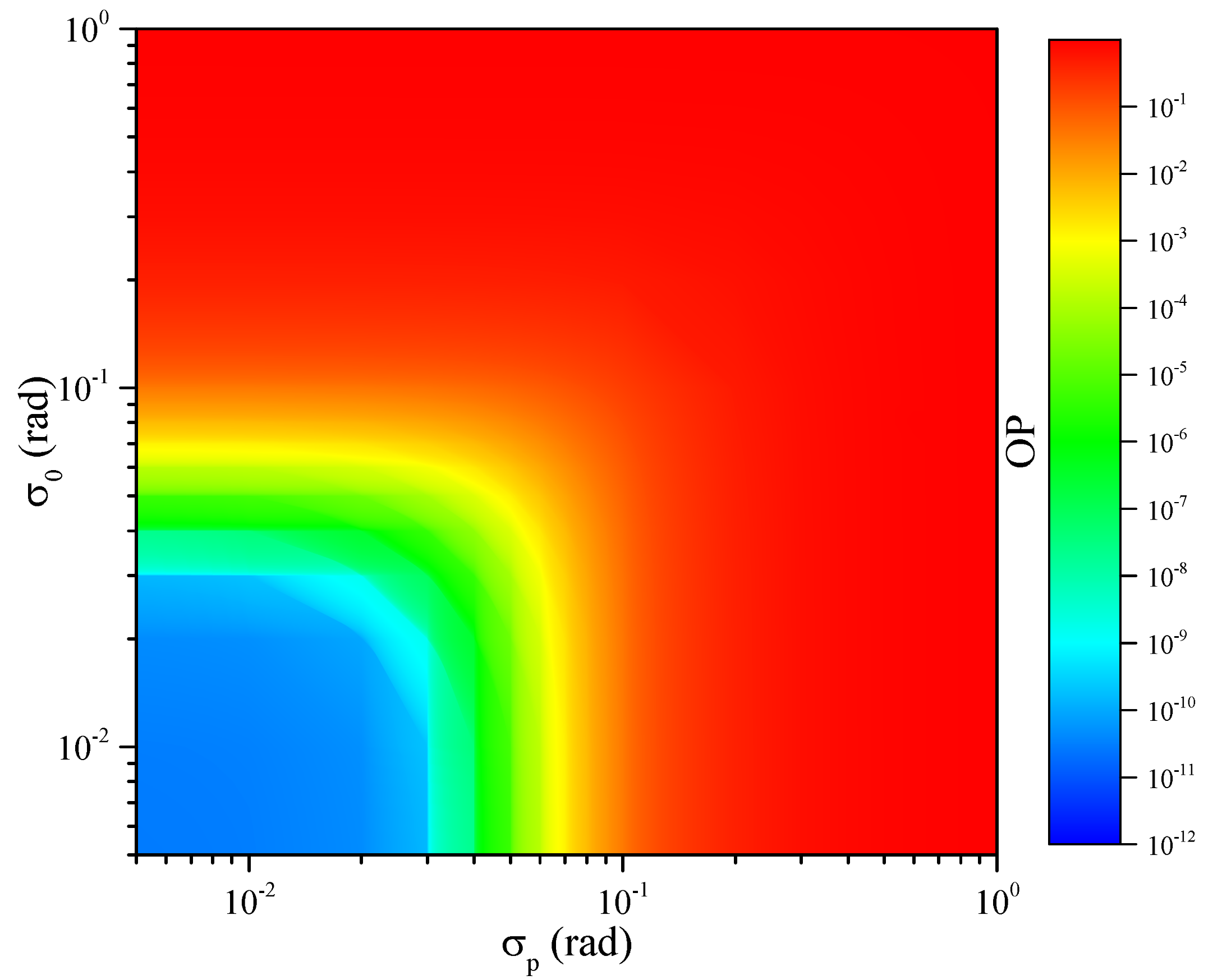}\\
	\vspace{-0.2cm}
	\centering (b)
	\end{minipage}
	\caption{OP vs $\sigma_p$ and $\sigma_0$ for a) $\mu=0.1\text{ }\mathrm{rad}$ and b) $\mu=1\text{ }\mathrm{rad}$.}
	\label{Fig:OP_vs_sp_s0}
	\vspace{-0.2cm}
\end{figure}  

Figure~\ref{Fig:OP_vs_sp_s0} illustrates the OP as a function of $\sigma_p$ and $\sigma_0$ for different values of $\mu$, assuming $N=16$, $K_r=5\text{ }\mathrm{dB}$, $m_1=1$, and $\gamma/\gamma_{\mathrm{th}}=5\text{ }\mathrm{dB}$. As expected, for given $\mu$ and $\sigma_0$, as $\sigma_p$ increases, the OP also increases. For example, for $\mu=0.1\text{ }\mathrm{rad}$ and $\sigma_0=0.1\text{ }\mathrm{rad}$, the OP shifts from $3.35\times 10^{-11}$ to $0.0364$, as $\sigma_p$ increases from $0.01$ to $0.1\text{ }\mathrm{rad}$. This example highlights the detrimental impact of disorientation and misalignment on the outage performance of RIS-assisted UAV wireless systems. Similarly, for fixed $\mu$ and $\sigma_p$, as $\sigma_0$ increases, the OP also increases. For instance, for $\mu=0.1\text{ }\mathrm{rad}$ and $\sigma_p=0.1\text{ }\mathrm{rad}$, as $\sigma_0$ increases from  $0.01$ to $0.1\text{ }\mathrm{rad}$, the OP changes from $0.035$ to $0.036$. On the other hand, for $\mu=1\text{ }\mathrm{rad}$ and $\sigma_p=0.1$, the same $\sigma_0$ shifts results to an OP change from $0.0364$ to $0.1845$. This reveals the importance of accounting $\mu$ when assessing the performance of RIS-assisted UAV wireless systems.         
    
\begin{figure}
	\centering\includegraphics[width=0.5\linewidth,trim=0 0 0 0,clip=false]{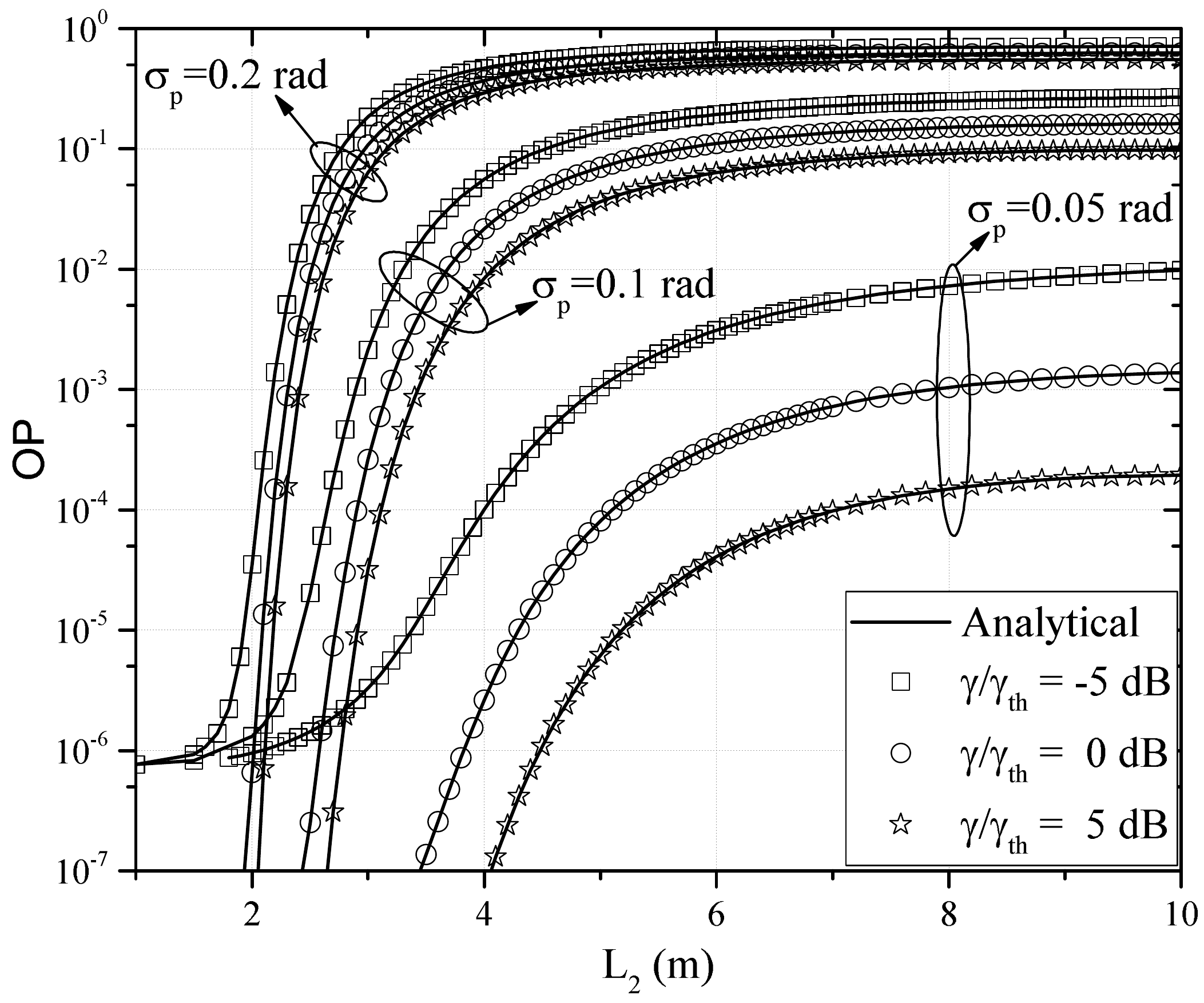}
	\vspace{-0.2cm}
	\caption{OP vs $L_2$ for different values of $\sigma_p$ and $\gamma/\gamma_{\mathrm{th}}$.}
	\label{Fig:OP_vs_L2_diff_sp_g}
	\vspace{-0.2cm}
\end{figure}
   
   Figure~\ref{Fig:OP_vs_L2_diff_sp_g} depicts the OP as a function of $L_2$, for different values of $\sigma_p$ and $\gamma/\gamma_{\mathrm{th}}$, assuming that $\sigma_0=d_x=0.1$, $m=1$, $K_r=5\text{ }\mathrm{dB}$ and $N=16$. As expected, for given $L_2$ and $\gamma/\gamma_{\mathrm{th}}$, as $\sigma_p$ increases, an outage performance degradation is observed. For example, for $L=8$ and $\gamma/\gamma_{\mathrm{th}}=5\text{ }\mathrm{dB}$, the OP increases for approximately $3$ orders of magnitude, as $\sigma_p$ increases from $0.05$ to $0.1\text{ }\mathrm{rad}$. Additionally, from this figure, we observe that for fixed $\sigma_p$ and $L$, the outage performance improves, as $\gamma/\gamma_{\mathrm{th}}$ increases. For instance, for  $\sigma_p=0.05\text{ }\mathrm{rad}$ and $L=10\text{ }\mathrm{m}$, the OP decreases from $9.8\times 10^{-3}$ to $1.9\times 10^{-4}$, as $\gamma/\gamma_{\mathrm{th}}$ increases from $-5$ to $5\text{ }\mathrm{dB}$, while, for the same $L$ and $\sigma_p=0.2\text{ }\mathrm{rad}$, the same $\gamma/\gamma_{\mathrm{th}}$ shift results to an OP change from $0.71$ to $0.55$. This indicates that, for a given $L_2$, as $\sigma_p$ increases, the same $\gamma/\gamma_{\mathrm{th}}$ variation results to a lower outage performance improvement. Additionally, for fixed $\sigma_p$ and $\gamma/\gamma_{\mathrm{th}}$, as $L_2$ increases, the equivalent beam-waist increases; thus, the OP also increases. For example, for $\sigma_p=0.1\text{ }\mathrm{rad}$ and $\gamma/\gamma_{\mathrm{th}}=0\text{ }\mathrm{dB}$, the OP increases for about $2$ orders of magnitude, as $L_2$ increases from $4$ to $8\text{ }\mathrm{m}$. Finally and most importantly, from this figure, it becomes apparent that there exists a relationship between, the maximum achievable transmission distance of the RIS-UAV hop for a give OP requirement, the level of disorientation and misalignment, the transmission power, which is directly connected to $\gamma$, and the spectral efficiency of the transmission scheme, $r_{\mathrm{th}}$ that is related to $\gamma_{\mathrm{th}}$ by $\gamma_{\mathrm{th}}=2^{r_{\mathrm{th}}}-1$. For instance, in the environment in which $\sigma_p=0.05\text{ }\mathrm{rad}$, and for an application that requires a maximum OP equal to $10^{-5}$, the maximum achievable transmission distance of the RIS-UAV link is $4\text{ }\mathrm{m}$, if $\gamma/\gamma_{\mathrm{th}}$ equals $-5\text{ }\mathrm{m}$.  
  
\subsection{Non-ideal RF front-end} \label{Eq:Non_ideal_RF_FE_results}

\subsubsection{Without disorientation and misalignment} 
This section is devoted to assess the joint impact of hardware imperfections and fading in RIS-assisted UAV wireless systems. In this direction, the effect of disorientation and misalignment is not~accounted.  

\begin{figure}
	\begin{minipage}{0.48\linewidth}
	\centering\includegraphics[width=1\linewidth,trim=0 0 0 0,clip=false]{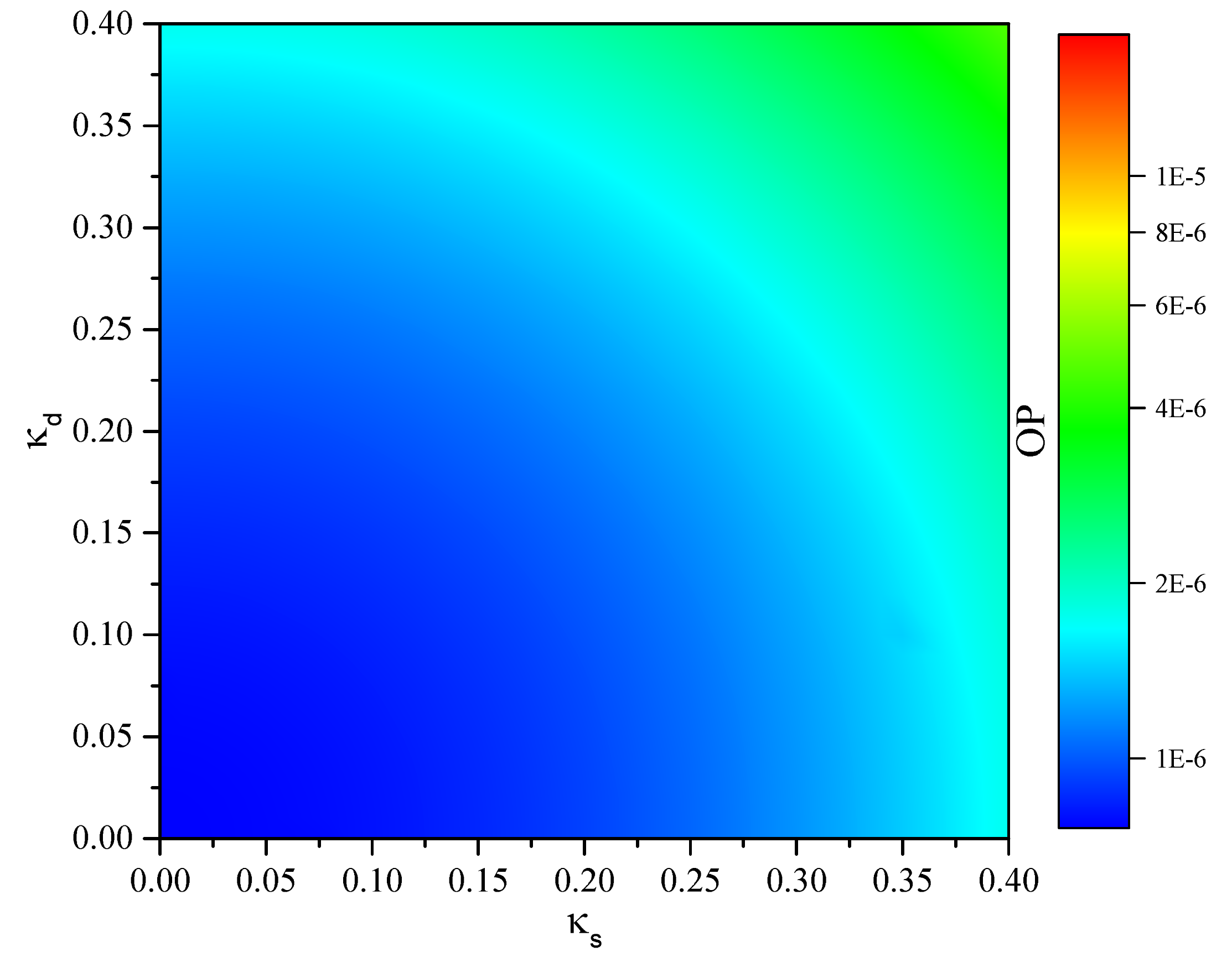}\\
	\vspace{-0.2cm}
	(a)
	\end{minipage}
	\begin{minipage}{0.48\linewidth}
	\centering\includegraphics[width=1\linewidth,trim=0 0 0 0,clip=false]{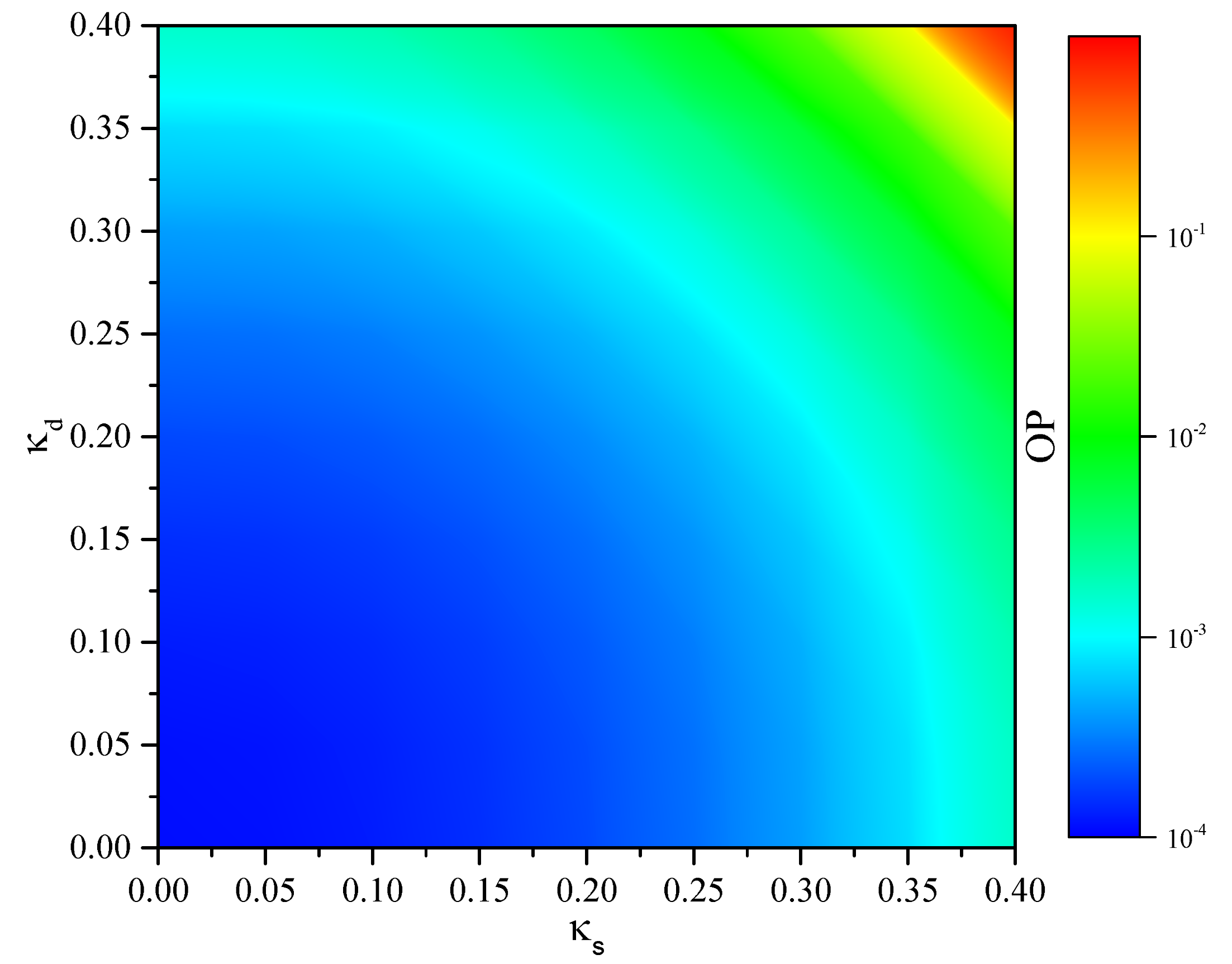}\\
	\vspace{-0.2cm}
	(b)
	\end{minipage}
	\centering\begin{minipage}{0.48\linewidth}
		\centering\includegraphics[width=1\linewidth,trim=0 0 0 0,clip=false]{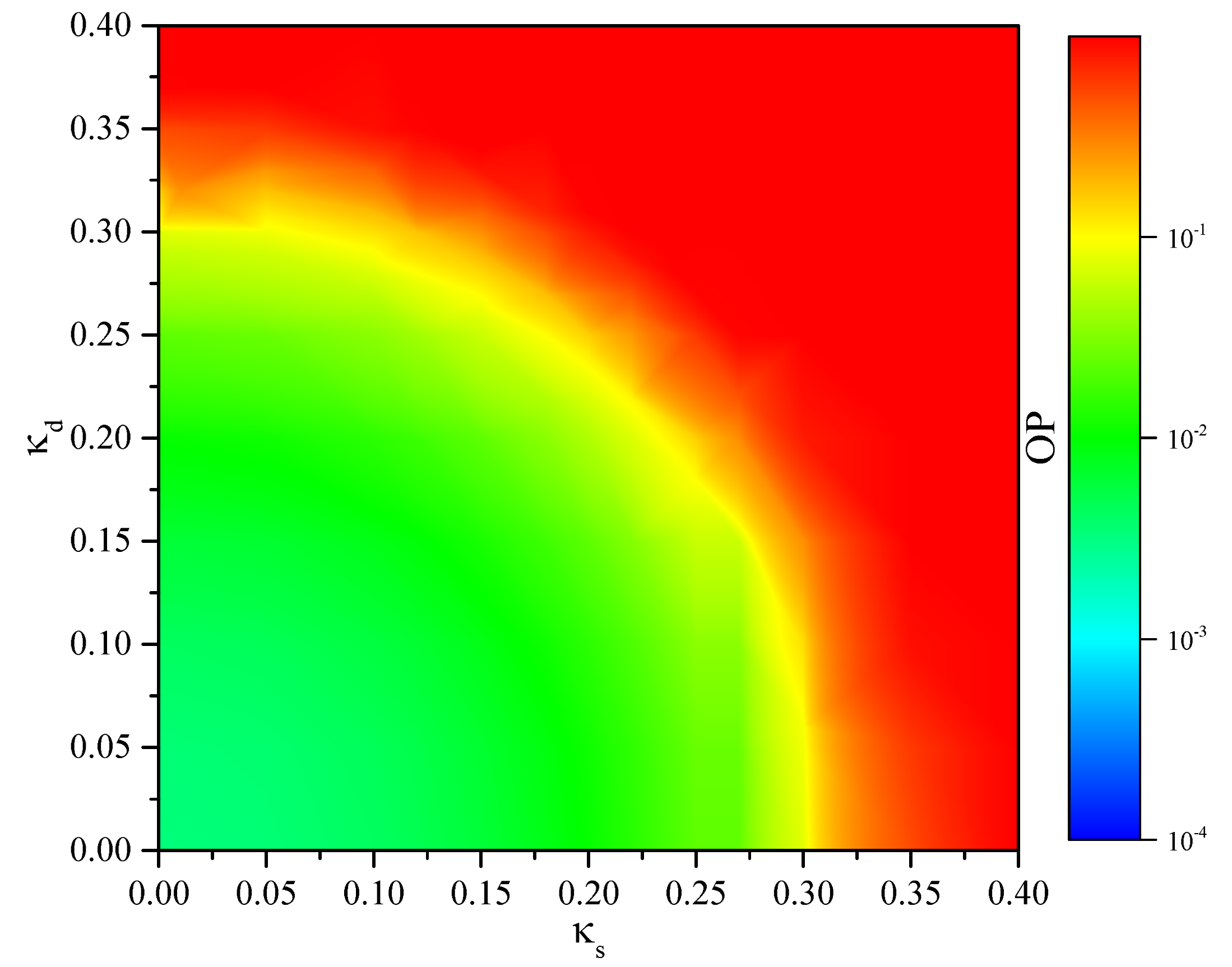}\\
	\vspace{-0.2cm}
	(c)
	\end{minipage}
	\vspace{-0.2cm}
	\caption{OP vs $\kappa_s$ and $\kappa_d$ for a) $\gamma_{\mathrm{th}}=1$, b) $\gamma_{\mathrm{th}}=3$, and c) $\gamma_{\mathrm{th}}=7$.}
	\label{Fig:OP_vs_ks_kd}
	\vspace{-0.2cm}
\end{figure} 

Figure~\ref{Fig:OP_vs_ks_kd} demonstrates the joint impact of hardware imperfections and fading on the outage performance of RIS-assisted UAV wireless systems. In particular, the OP is plotted as a function of $\kappa_s$ and $\kappa_d$, for different values of $\gamma_{\mathrm{th}}$, assuming $m=1$, $K_r=5\text{ }\mathrm{dB}$, $N=16$, and~$\gamma=-5\text{ }\mathrm{dB}$. Notice that $\kappa_s=0$ indicates that the S is equipped with ideal RF front-end. Similarly, $\kappa_d=0$ means that the UAV is equipped with ideal RF front-end. Hence, the $\left(\kappa_{s}, \kappa_d\right)=(0,0)$ point represents the case in which both the S transmitter and UAV receiver are ideal. As expected, for given $\gamma_{\mathrm{th}}$ and $\kappa_s$, as $\kappa_d$ increases, the OP also increases. For example, for  $\gamma_{\mathrm{th}}=1$ and $\kappa_s=0.1$, the OP doubles, as $\kappa_d$ increases from $0.2$ to $0.4$. Similarly, for fixed $\gamma_{\mathrm{th}}$ and $\kappa_d$, as $\kappa_s$ increases the outage performance degrades. Moreover, it is worth-noting that, for a given $\gamma_{th}$, the same outage performance degradation happens for the following two scenarios: i) $\kappa_s=u_1$ and $\kappa_d$ changes from $u_2$ to $u_3$, and ii) $\kappa_d=c_1$ and $\kappa_s$ changes from $c_2$ to $u_3$, where $u_1, u_2$, $u_3$ are constants and $u_2>u_3$. Additionally, from this figure it becomes evident that for a given set of $\kappa_s$ and $\kappa_d$, as $\gamma_{\mathrm{th}}$ increases, i.e. the spectral efficiency of the transmission signal increases, the OP also increases. Finally, for $\gamma_{\mathrm{th}}\geq \frac{1}{\sqrt{\kappa_s^2 + \kappa_d^2}}$, the OP becomes equal to $1$.   

\begin{figure}
	\centering\includegraphics[width=0.5\linewidth,trim=0 0 0 0,clip=false]{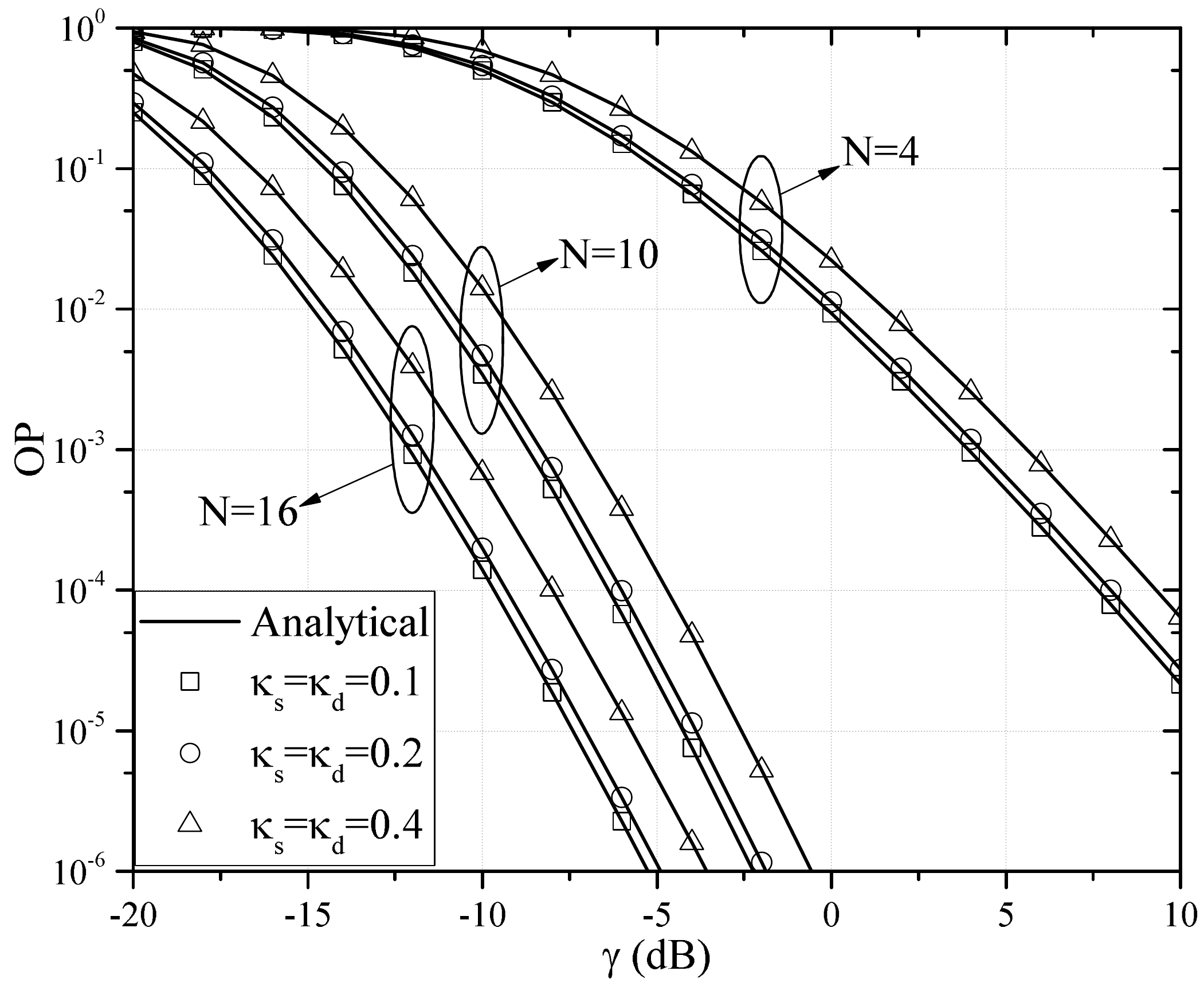}
	\vspace{-0.2cm}
	\caption{OP vs $\gamma$, for different values of $\kappa_s$, $\kappa_d$ and $N$.}
	\label{Fig:OP_vs_gamma_diff_ks_kd_N}
	\vspace{-0.2cm}
\end{figure}   

Figure~\ref{Fig:OP_vs_gamma_diff_ks_kd_N} depicts the OP as a function of $\gamma$, for different values of $\kappa_s$, $\kappa_d$ and $N$, assuming $m=1$, $K_r=5\text{ }\mathrm{dB}$, and $\gamma_{\mathrm{th}}=1$. As expected, for fixed $\kappa_s$, $\kappa_d$ and $N$, as $\gamma$ increases, the outage performance improves. Likewise, for given $\gamma$, and $N$, as the level of hardware imperfections increases, the OP also increases. For example, for $\gamma=-5\text{ }\mathrm{dB}$ and $N=10$, the OP increases by approximately one order of magnitude, as $\kappa_s=\kappa_d$ changes from $0.1$ to $0.4$. Finally, we observe that the impact of hardware imperfections on the outage performance becomes more severe as $N$ increases. In more detail, for a give OP requirement and a fixed $\kappa_s=\kappa_d$ change, the required $\gamma$ shift increases as $N$ increases. For instance, for a required OP of $10^{-4}$,  a $\kappa_s=\kappa_d$ change from $0.1$ to $0.4$, and $N=16$, $\gamma$ should be increased by approximately $2\text{ }\mathrm{dB}$. On the other hand, for the same requirement and $\kappa_s=\kappa_d$ change, the $\gamma$ should be increased by approximately $1\text{ }\mathrm{dB}$, for the case in which $N=4$.

\subsubsection{With disorientation and misalignment}  
In this section, the joint impact of fading, disorientation, misalignment, and transceivers' hardware imperfections is graphically illustrated. Unless otherwise stated, we assume that $\sigma_p=0.05\text{ }\mathrm{rad}$, $\sigma_0=0.1\text{ }\mathrm{rad}$ and $d_x=0.1$. 

\begin{figure}
	\centering\includegraphics[width=0.5\linewidth,trim=0 0 0 0,clip=false]{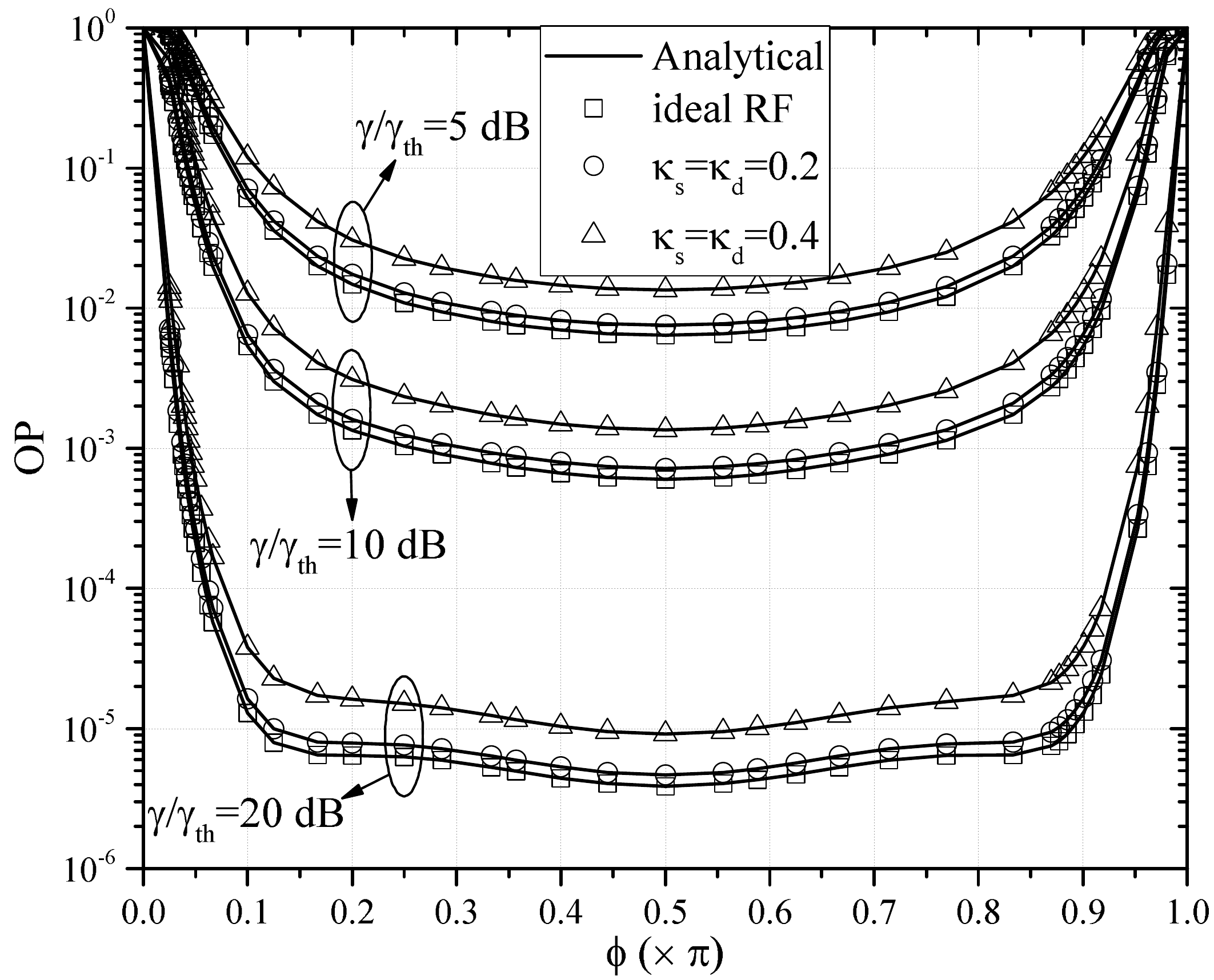}
	\vspace{-0.2cm}
	\caption{OP vs $\phi$, for different values of $\kappa_s$, $\kappa_d$ and $\gamma/\gamma_{\mathrm{th}}$.}
	\label{Fig:OP_vs_phi}
	\vspace{-0.2cm}
\end{figure} 

In Fig.~\ref{Fig:OP_vs_phi},  the OP is plotted as a function of $\phi$, for different values of $\gamma/\gamma_{\mathrm{th}}$, $\kappa_s$, and $\kappa_d$, assuming $N=16$, $m=1$, and $K_r=5\text{ }\mathrm{dB}$. As a benchmark, the case in which both the S transmitter and UAV receiver are equipped with ideal RF front-ends. For given $\phi$ and $\gamma/\gamma_{\mathrm{th}}$, as the levels of RF imperfections increases, the outage performance degrades. Moreover, for fixed $\phi$, $\kappa_s$ and $\kappa_d$, as $\gamma/\gamma_{\mathrm{th}}$, the OP decreases. Likewise, for given $\gamma/\gamma_{\mathrm{th}}$, $\kappa_s$, $\kappa_d$ and $\phi\in\left[0,\frac{\pi}{2}\right)$, as $\phi$ increases, the OP decreases, while, for $\phi\in\left(\frac{\pi}{2}, \pi\right]$, as $\phi$ increases, the OP increases. The minimum OP is observed for $\phi=\frac{\pi}{2}$, i.e., when the expected value of the UAV center and the position of the RIS center have the same elevation. Finally, we observe that for fixed $\gamma/\gamma_{\mathrm{th}}$, $\kappa_s$, and $\kappa_d$, the same OP is achieved for $\phi$ and $\frac{\pi}{2}-\phi$. 

\begin{figure}
	\centering\includegraphics[width=0.5\linewidth,trim=0 0 0 0,clip=false]{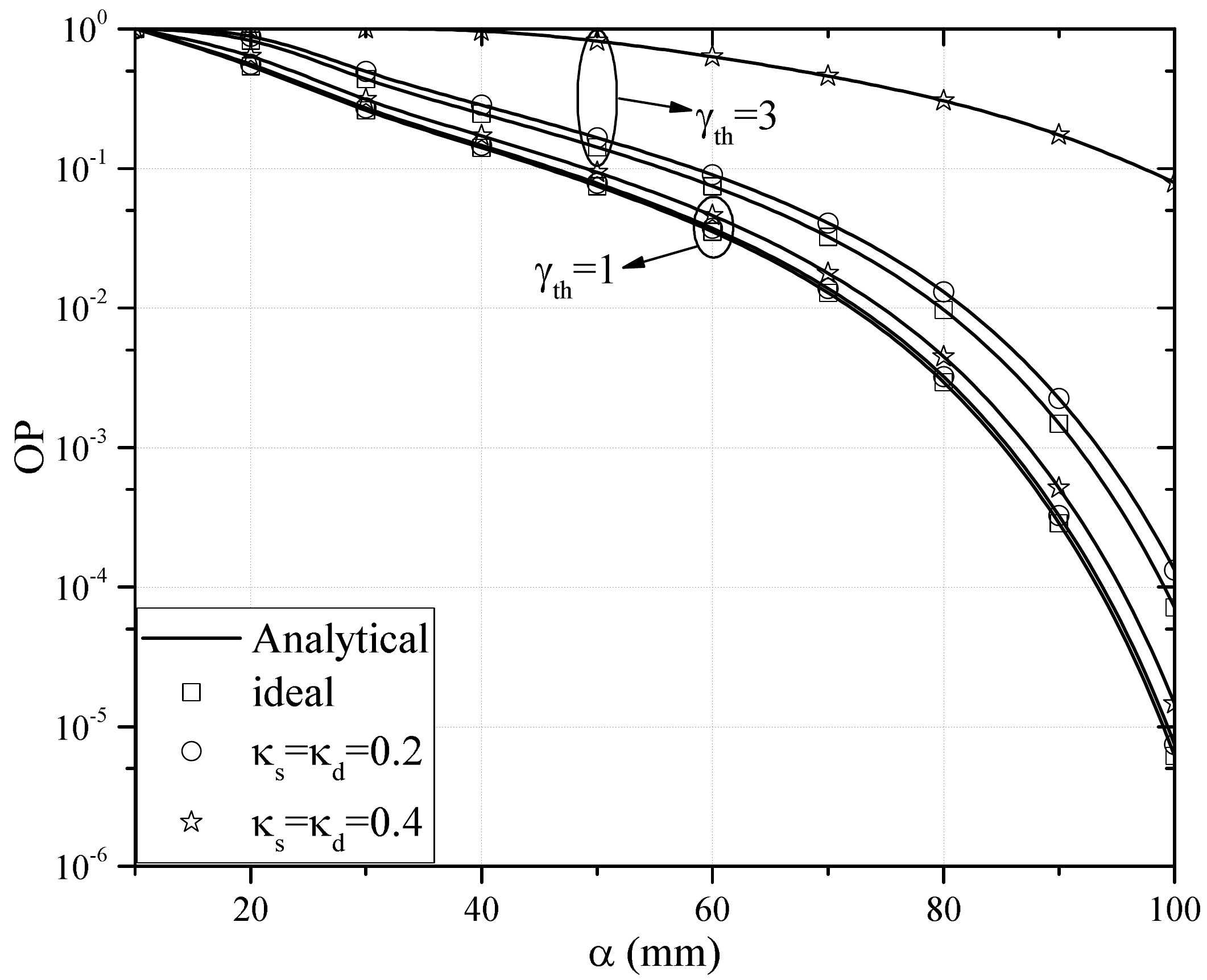}
	\vspace{-0.2cm}
	\caption{OP vs $\alpha$, for different values of $\kappa_s$, $\kappa_d$ and $\gamma_{\mathrm{th}}$.}
	\label{Fig:OP_vs_alpha}
	\vspace{-0.2cm}
\end{figure}               

Figure~\ref{Fig:OP_vs_alpha} depicts the OP as a functions of the UAV receiver effective area, $\alpha$, for different values of $\kappa_s$, $\kappa_d$ and $\gamma_{\mathrm{th}}$, assuming $m=1$, $K_r=5\text{ }\mathrm{dB}$, and $\gamma=5\text{ }\mathrm{dB}$. For given $\kappa_s$, $\kappa_d$, and $\gamma_{\mathrm{th}}$, as $\alpha$ increases, the portion of the RIS beam footprint at the UAV reception plane that falls outside the effective area decreases; as a result, the OP also decreases. For example, for $\kappa_s=\kappa_d=0.2$, and $\gamma_{\mathrm{th}}=1$, the OP decreases form more than four orders of magnitude, as $\alpha$ increases from $40$ to $100\text{ }\mathrm{mm}$. Additionally, for given $\alpha$ and $\gamma_{\mathrm{th}}$, as the level of transceivers hardware imperfections increases, the OP also increases. For instance, for $\alpha=50\text{ }\mathrm{mm}$ and $\gamma_{\mathrm{th}}=1$, the OP increases from $0.075$ to $0.093$, as $\kappa_s=\kappa_d$ increases from $0$ to $0.4$, whereas, for the same $\alpha$ and $\kappa_s=\kappa_d$ increase as well as $\gamma_{\mathrm{th}}=3$, the OP increases from  $0.14$ to $0.81$. This indicates that as $\gamma_{\mathrm{th}}$ increases, the impact of hardware imperfections become more severe.        

\vspace{-0.5cm}
\section{Conclusions}\label{S:Conclusions} \vspace{-0.2cm}

In this paper, we presented a theoretical framework for quantifying the outage performance of RIS-assisted UAV wireless systems that takes into account the effects of various types of fading, disorientation and misalignment, as well as transceiver hardware imperfections. In more detail, we stochastically characterized the e2e channel by extracting closed-form expressions for its PDF and CDF. Based on this, we derived closed-form formulas for the OP and diversity order for the following cases: (i) both the S-transmitter and UAV-receiver are equipped with ideal RF front-end and the RIS-UAV link experience neither disorientation nor misalignment, (ii) both the S-transmitter and UAV-receiver are equipped with ideal RF front-end and the RIS-UAV link experience disorientation and misalignment, (iii) both the S-transmitter and UAV-receiver are equipped with non-ideal RF front-end and the RIS-UAV link experience neither disorientation nor misalignment, and (iv) both the S transmitter and UAV receiver are equipped with non-ideal RF front-end and the RIS-UAV link experience disorientation and misalignment. Simulation results were used in order to validate the theoretical framework and demonstrated the importance of accurately modeling small-scale fading conditions, disorientation and misalignment as well as transceiver hardware imperfections.       

\vspace{-0.3cm}
\section*{Appendices} \vspace{-0.2cm}
\section*{Appendix A}  \vspace{-0.2cm}
\section*{Proof of Theorem~\ref{L:PDF_CDF_A}}

From~\eqref{Eq:A_final}, we can rewrite $A$~as \vspace{-0.2cm}
\begin{align}
	A = \sum_{i=1}^{N}\chi_i,
	\label{Eq:A_final_2}
\end{align}
where \vspace{-0.3cm}
\begin{align}
	\chi_i=\left|h_i\right| \left|g_i\right|.
	\label{Eq:chi_i}
\end{align}
Since $\left|h_i\right|$ and $\left|g_i\right|$, the PDF of $\chi_i$ can be analytically evaluated~as \vspace{-0.2cm}
\begin{align}
	f_{\chi_i}(x) = \int_{0}^{\infty} \frac{1}{y} f_{h_i}(y) f_{g_i}\left(\frac{x}{y}\right)\,\mathrm{d}y,
\end{align}
which, by applying~\eqref{Eq:f_h_i} and~\eqref{Eq:f_g_i} can be equivalently written~as \vspace{-0.2cm} 
\begin{align}
	f_{\chi_i}(x) = \sum_{m=1}^{M} \sum_{k=1}^{K} 4 a_{m}^{(1)} a_k^{(2)} x^{2 b_{k}^{(2)}-1} \mathcal{I}_{m,k}\left(x\right),
	\label{Eq:f_chi_i}
\end{align}
where \vspace{-0.3cm}
\begin{align}
	\mathcal{I}_{m,k}\hspace{-0.1cm}\left(x\right) \hspace{-0.1cm} = \hspace{-0.1cm} \int_0^{\infty} \hspace{-0.1cm} y^{2 b_m^{(1)} - 2 b_k^{(2)}-1} \exp\left(-c_1 y^2 - \frac{c_2 x^2}{y^2}\right)\,\mathrm{d}y. 
	\label{Eq:I}
\end{align}
With the aid of~\cite[Eq. (3.478/4)]{B:Gra_Ryz_Book},~\eqref{Eq:I} can be expressed in closed-form~as \vspace{-0.2cm}
\begin{align}
	\mathcal{I}_{m,k}\hspace{-0.1cm}\left(x\right)\hspace{-0.1cm} =\hspace{-0.1cm} \left(\frac{c_1}{c_2}\right)^{-\frac{ b_m^{(1)}- b_k^{(2)}}{2}} x^{{b_m^{(1)}}-b_k^{(2)}} \mathrm{K}_{{b_m^{(1)}}-{b_k^{(2)}}}\left(2\sqrt{c_1 c_2} x\right).
	\label{Eq:I_s2}
\end{align}
By~applying~\eqref{Eq:I_s2} into~\eqref{Eq:f_chi_i}, we get \vspace{-0.2cm}
\begin{align}
	f_{\chi_i}(x) = \sum_{m=1}^{M}  \sum_{k=1}^{K} & 4 a_{m}^{(1)} a_k^{(2)} \left(\frac{c_1}{c_2}\right)^{-\frac{ b_m^{(1)}- b_n^{(2)}}{2}} x^{{b_m^{(1)}}+b_k^{(2)}-1}
	\nonumber \\ & \times
	 \mathrm{K}_{{b_m^{(1)}}-{b_k^{(2)}}}\left(2\sqrt{c_1 c_2} x\right),
	 \label{Eq:f_chi_i_semifinal}
\end{align}
or equivalently \vspace{-0.2cm}
\begin{align}
	f_{\chi_i}(x) &= \sum_{m=1}^{M}   \sum_{k=1}^{K} a_m^{(1)} a_k^{(2)}
	\left(\frac{c_1}{c_2}\right)^{-\frac{b_m^{(1)}-b_k^{(2)}}{2}} \left(c_1 c_2\right)^{-\frac{b_m^{(1)}+b_k^{(2)}}{2}}
	\nonumber \\ & \times 
	\Gamma\left({b_m^{(1)}}\right)
	\Gamma\left(b_k^{(2)}\right) 
	f_{\mathrm{K}_G}^{(m,k)}\left(x\right),
	\label{Eq:f_chi_i_final}
\end{align}
where \vspace{-0.2cm}
\begin{align}
	f_{\mathrm{K}_G}^{(m,k)}\left(x\right) &= 
	\frac{4\left(c_1 c_2\right)^{\frac{b_m^{(1)}+b_k^{(2)}}{2}}}{\Gamma\left(b_m^{(1)}\right)\Gamma\left(b_k^{(2)}\right)} 
	x^{b_m^{(1)}+b_k^{(2)}-1}
	\nonumber \\ & \times 
	\mathrm{K}_{b_m^{(1)}-b_k^{(2)}}\left(2\sqrt{c_1 c_2} x\right). 
	\label{Eq:f_K_G}
\end{align}
Notice that~\eqref{Eq:f_K_G} is a special case of the PDF of the generalized-K distribution. Thus, from~\eqref{Eq:f_chi_i_final}, it becomes apparent that $\chi_i$ follows a mixture generalized-K distribution. Likewise, from~\eqref{Eq:A_final_2}, we observe that $A$ is a sum of $N$ mixture generalized-K distributed RVs; hence, according to~\cite[Eqs. (6) and (7)]{Peppas2011}, its PDF and CDF can be respectively obtained as in~\eqref{Eq:f_A} and ~\eqref{Eq:F_A}. Similarly,~\eqref{Eq:k_A}--\eqref{Eq:mu_A} can be extracted by applying~\cite[Eq. (17)]{Peppas2011}. Finally, $\mu_{\chi_i}(l)$ can be analytically evaluated~as \vspace{-0.2cm}
\begin{align}
	\mu_{\chi_i}(l) = \int_0^{\infty} x^{n} f_{\chi_i}(x)\,\mathrm{d}x,
\end{align}
which, by applying~\eqref{Eq:f_chi_i_semifinal}, can be rewritten~as \vspace{-0.2cm}
\begin{align}
	\mu_{\chi_i}(l) = \sum_{m=1}^{M}  \sum_{k=1}^{K} & 4 a_{m}^{(1)} a_k^{(2)} \left(\frac{c_1}{c_2}\right)^{-\frac{ b_m^{(1)}- b_n^{(2)}}{2}} \mathcal{K}_{m,n}, 
	\label{Eq:mu_chi_i_l_s0}
\end{align}
where \vspace{-0.2cm}
\begin{align}
	\mathcal{K}_{m,n} = \int_0^{\infty}   x^{{b_m^{(1)}}+b_k^{(2)}+n-1}
	\mathrm{K}_{{b_m^{(1)}}-{b_k^{(2)}}}\left(2\sqrt{c_1 c_2} x\right)\,\mathrm{d}x.
	\label{Eq:K}
\end{align}
By using~\cite[Eq. (6.561/16)]{B:Gra_Ryz_Book} in~\eqref{Eq:K}, we obtain \vspace{-0.2cm}
\begin{align}
	\mathcal{K}_{m,n} = \frac{1}{4} \left(c_1 c_2\right)^{-\frac{{b_m^{(1)}}+b_k^{(2)}+n}{2}}\,\Gamma\left({b_m^{(1)}}+\frac{n}{2} \right)\,\Gamma\left({b_k^{(2)}}+\frac{n}{2} \right).
	\label{Eq:K_s2}
\end{align}
Finally, with the of~\eqref{Eq:K_s2}, \eqref{Eq:mu_chi_i_l_s0} can be written as~\eqref{Eq:mu_x}. 
This concludes the proof.  

\vspace{-0.2cm}
\section*{Appendix B}
\section*{Proof of Theorem~\ref{Th:PDF_CDF_Ae2e}} 

Since $h_g$ and $A$ are independent RVs, the CDF of $A_{e2e}$ can be evaluated~as \vspace{-0.2cm}
\begin{align}
	F_{A_{e2e}}(x) = \int_{0}^{B_o} F_{A}\left(\frac{x}{y}\right) f_{h_g}(y)\,\mathrm{d}y. 
	\label{Eq:F_A_e2e}
\end{align}
By employing~\eqref{Eq:f_h_g} and~\eqref{Eq:F_A},~\eqref{Eq:F_A_e2e} can be expressed~as \vspace{-0.2cm}
\begin{align}
	F_{A_{e2e}}(x) =  \frac{\rho}{B_o^{\zeta}}\frac{1}{\Gamma\left(k_A\right)\Gamma\left(m_A\right)} \mathcal{J}(x),
	\label{Eq:F_A_e2e_appendix}
\end{align}
where \vspace{-0.2cm}
\begin{align}
	\mathcal{J}(x) = \int_{0}^{B_o} y^{\zeta-1} \mathrm{G}_{1, 3}^{2, 1}\left(\Xi^2 \frac{x^2}{y^2}\left| \begin{array}{c} 1 \\ k_A, m_A, 0 \end{array} \right. \right) \, \mathrm{d}y. 
	\label{Eq:J}
\end{align}
With the aid of~\cite[Eq. (07.34.17.0012.01)]{WS:mathematica_function},~\eqref{Eq:J} can be equivalently written as \vspace{-0.2cm}
\begin{align}
	\mathcal{J}(x) = \int_{0}^{B_o} y^{\zeta-1} \mathrm{G}_{ 3, 1}^{ 1, 2}\left(\Xi^{-2} \frac{y^2}{x^2}\left| \begin{array}{c}  1-k_A, 1-m_A, 1 \\ 0 \end{array} \right. \right) \, \mathrm{d}y,
	\label{Eq:Js2}
\end{align}
which, by employing~\cite{W:07.34.21.0084.01}, yields \vspace{-0.2cm}
\begin{align}
	\mathcal{J}(x) = \frac{B_o^{\zeta}}{2\zeta} \mathrm{G}_{5,3}^{1,4}\left(\left.\begin{array}{c}1-k_A, 1-m_A, \frac{1-\zeta}{2}, \frac{2-\zeta}{2},1 \\ 0, \frac{1-\zeta}{2}, - \frac{\zeta}{2}\end{array}\right| \frac{B_o^2}{\Xi^2 x^2}\right).
	\label{Eq:Js3}
\end{align}
By applying~\eqref{Eq:Js3} into~\eqref{Eq:F_A_e2e_appendix}, we obtain~\eqref{Eq:F_A_e2e_final_general}. 

The PDF of $A_{e2e}$ can be evaluated~as
%\begin{align}
$	f_{A_{e2e}}\left(x\right) = \frac{\mathrm{d}F_{A_{e2e}}(x)}{\mathrm{d}x}, $
%\end{align}
which, by employing~\eqref{Eq:F_A_e2e_final_general}, can be rewritten~as \vspace{-0.2cm}
\begin{align}
	f_{A_{e2e}}&\left(x\right) = \frac{\zeta}{2\Gamma\left(k_A\right)\Gamma\left(m_A\right)}
	\nonumber \\ & \hspace{-1cm}\times \frac{\mathrm{d}\mathrm{G}_{5,3}^{1,4}\left(\left.\begin{array}{c}1-k_A, 1-m_A, \frac{1-\zeta}{2}, \frac{2-\zeta}{2},1 \\ 0, \frac{1-\zeta}{2}, - \frac{\zeta}{2}\end{array}\right| \frac{B_o^2}{\Xi^2 x^2}\right)}{\mathrm{d}x},
\end{align} 
or, as in~\eqref{Eq:f_A_e2e_final}. This concludes the proof. 

\vspace{-0.2cm}
\section*{Appendix C} \vspace{-0.1cm}
\section*{Proof of Proposition 4} \vspace{-0.1cm}

The OP can be expressed~as \vspace{-0.2cm}
\begin{align}
	P_o^{\text{wo}}\left(\gamma_{\mathrm{th}}\right) = \Pr\left(\gamma_u\leq\gamma_{\mathrm{th}}\right),
	\label{Eq:P_o_wo_s1}
\end{align}
which, with the aid of~\eqref{Eq:gamma_u}, can be rewritten~as \vspace{-0.2cm}
\begin{align}
	P_o^{\text{wo}}\left(\gamma_{\mathrm{th}}\right) =\Pr\left(\frac{A^2}{\left(\kappa_s^2+\kappa_d^2\right) A^2 + \frac{1}{\gamma}} \leq \gamma_{\mathrm{th}} \right), 
\end{align}
or equivalently \vspace{-0.2cm}
\begin{align}
	P_o^{\text{wo}}\left(\gamma_{\mathrm{th}}\right) =\Pr\left( A^2 \left(1-\left(\kappa_s^2 + \kappa_d^2\right) \gamma_{\mathrm{th}}\right)\leq \frac{\gamma_{\mathrm{th}}}{\gamma} \right).
	\label{Eq:P_o_wo_s2}
\end{align}

For~$\left(1-\left(\kappa_s^2 + \kappa_d^2\right) \gamma_{\mathrm{th}}\right) \leq 0$, the condition  $A^2 \left(1-\left(\kappa_s^2 + \kappa_d^2\right) \gamma_{\mathrm{th}}\right)\leq \frac{\gamma_{\mathrm{th}}}{\gamma}$ is always valid; thus, the OP equals $1$, i.e., \vspace{-0.2cm} 
\begin{align}
	P_o^{\text{wo}}\left(\gamma_{\mathrm{th}}\right) = 1, \text{ for } \gamma_{\mathrm{th}}\geq \frac{1}{\kappa_s^2 + \kappa_d^2}.
	\label{Eq:P_o_wo_case1}
\end{align}
On the other hand, for~$\left(1-\left(\kappa_s^2 + \kappa_d^2\right) \gamma_{\mathrm{th}}\right) > 0$,~\eqref{Eq:P_o_wo_s2} can be written~as \vspace{-0.2cm}
\begin{align}
	P_o^{\text{wo}}\left(\gamma_{\mathrm{th}}\right) &= \Pr\left(A \leq \frac{1}{\sqrt{1-\left(\kappa_s^2+\kappa_d^2\right)\gamma_{\mathrm{th}}}} \sqrt{\frac{\gamma_{\mathrm{th}}}{\gamma}}\right), \nonumber \\  &\hspace{+3cm}\text{ for } \gamma_{\mathrm{th}}< \frac{1}{\kappa_s^2 + \kappa_d^2},
\end{align}  
or \vspace{-0.3cm}
\begin{align}
	P_o^{\text{wo}}\left(\gamma_{\mathrm{th}}\right) &= F_A\left( \frac{1}{\sqrt{1-\left(\kappa_s^2+\kappa_d^2\right)\gamma_{\mathrm{th}}}} \sqrt{\frac{\gamma_{\mathrm{th}}}{\gamma}}\right), \nonumber \\  &  \hspace{+3cm} \text{ for } \gamma_{\mathrm{th}}< \frac{1}{\kappa_s^2 + \kappa_d^2},
\end{align}
which, by applying~\eqref{Eq:F_A}, yields \vspace{-0.2cm}
\begin{align}
	P_o^{\text{wo}}\left(\gamma_{\mathrm{th}}\right) &= \frac{\mathrm{G}_{1, 3}^{2, 1}\left( \frac{\Xi^2}{1-\left(\kappa_s^2+\kappa_d^2\right)\gamma_{\mathrm{th}}} \frac{\gamma_{\mathrm{th}}}{\gamma}\left|\begin{array}{c} 1\\ k_A, m_A, 0 \end{array}\right.\right)}{\Gamma\left(k_A\right)\Gamma\left(m_A\right)}, \label{Eq:P_o_wo_case2}  \\  & \hspace{+4cm}\text{ for } \gamma_{\mathrm{th}}< \frac{1}{\kappa_s^2 + \kappa_d^2}.\nonumber 
\end{align}
Finally, by combining~\eqref{Eq:P_o_wo_case1} and~\eqref{Eq:P_o_wo_case2}, we obtain~\eqref{Eq:Po_wo_final}. 
This concludes the proof.

\vspace{-0.5cm}
\section*{Appendix D}
\section*{Proof of Proposition 5}
From~\eqref{Eq:P_o_wo_s1}, in the presence of disorientation and misalignment, the OP can be expressed~as \vspace{-0.2cm}
\begin{align}
	P_o^{w}\left(\gamma_{\mathrm{th}}\right)=\Pr\left(A_{e2e}^2\left(1-\left(\kappa_s^2 + \kappa_d^2\right) \gamma_{\mathrm{th}}\right)\leq \frac{\gamma_{\mathrm{th}}}{\gamma}\right).  
	\label{Eq:P_o_w_s1}
\end{align} 
For the case in which $\left(1-\left(\kappa_s^2 + \kappa_d^2\right) \gamma_{\mathrm{th}}\right) \leq 0$, the condition $A_{e2e}^2\left(1-\left(\kappa_s^2 + \kappa_d^2\right) \gamma_{\mathrm{th}}\right)\leq \frac{\gamma_{\mathrm{th}}}{\gamma}$ always holds; as a result, \vspace{-0.2cm}
\begin{align}
	P_o^{w}\left(\gamma_{\mathrm{th}}\right)= 1, \text{ for } \gamma_{\mathrm{th}}\geq \frac{1}{\kappa_s^2 + \kappa_d^2}. 
	\label{Eq:P_o_w_s3}
\end{align}
On the other hand, for $\left(1-\left(\kappa_s^2 + \kappa_d^2\right) \gamma_{\mathrm{th}}\right) > 0$,~\eqref{Eq:P_o_w_s1} can be equivalently written~as \vspace{-0.2cm}
\begin{align}
	P_o^{\text{w}}\left(\gamma_{\mathrm{th}}\right) &= \Pr\left(A_{e2e} \leq \frac{1}{\sqrt{1-\left(\kappa_s^2+\kappa_d^2\right)\gamma_{\mathrm{th}}}} \sqrt{\frac{\gamma_{\mathrm{th}}}{\gamma}}\right),  \nonumber \\  &\hspace{+2.8cm}\text{ for } \gamma_{\mathrm{th}}< \frac{1}{\kappa_s^2 + \kappa_d^2},
\end{align}
or \vspace{-0.3cm}
\begin{align}
	P_o^{\text{w}}\left(\gamma_{\mathrm{th}}\right) &= F_{A_{e2e}}\left(\frac{1}{\sqrt{1-\left(\kappa_s^2+\kappa_d^2\right)\gamma_{\mathrm{th}}}} \sqrt{\frac{\gamma_{\mathrm{th}}}{\gamma}}\right),\label{Eq:P_o_w_s2} \nonumber \\  &\hspace{+2.8cm}\text{ for } \gamma_{\mathrm{th}}< \frac{1}{\kappa_s^2 + \kappa_d^2}. 
\end{align}
By applying~\eqref{Eq:F_A_e2e_final_general} into~\eqref{Eq:P_o_w_s2}, we get \vspace{-0.2cm}
\begin{align}
	P_{o}^{\text{w}}(\gamma_{\mathrm{th}})&=
		\frac{\zeta \mathrm{G}_{5,3}^{1,4}\left(\left.\begin{array}{c}1-k_A, 1-m_A, \frac{1-\zeta}{2}, \frac{2-\zeta}{2},1 \\ 0, \frac{1-\zeta}{2}, - \frac{\zeta}{2}\end{array}\right| \frac{B_o^2}{\Xi^2} \frac{\gamma}{\gamma_{\mathrm{th}}}\right)}{2\Gamma\left(k_A\right)\Gamma\left(m_A\right)}, \label{Eq:P_o_w_s4} \nonumber \\ & \hspace{+3.8cm} \text{ for } \gamma_{\mathrm{th}}< \frac{1}{\kappa_s^2 + \kappa_d^2}  
\end{align}
Finally, by combining~\eqref{Eq:P_o_w_s3} and~\eqref{Eq:P_o_w_s4}, we obtain~\eqref{Eq:P_o_w}. This concludes the proof. 

\vspace{-0.5cm}
\balance
\bibliographystyle{IEEEtran}
\bibliography{IEEEabrv,References}

\end{document}